\documentclass[twocolumn,english]{revtex4-2}
\usepackage[utf8]{inputenc}
\usepackage[T1]{fontenc}
\usepackage{amsmath}
\usepackage{amsfonts}
\usepackage{amssymb}
\usepackage[version=4]{mhchem}
\usepackage{stmaryrd}
\usepackage{hyperref}
\hypersetup{colorlinks=true, linkcolor=blue, filecolor=magenta, urlcolor=cyan,}
\usepackage{mathrsfs}
\usepackage[dvips]{graphicx}
\usepackage{bm}
\usepackage{amsthm,amssymb,amsmath,amsfonts,latexsym,enumerate,float}
\usepackage[utf8]{inputenc}
\usepackage{cmap}

\begin{document}
\title{Structure-property relationships for weak-ferromagnetic perovskites }

\author{A.S. Moskvin}
\affiliation{Ural Federal University, 620083 Ekaterinburg, Russia}
\affiliation{M.N. Mikheev lnstitute of Metal Physics of Ural Branch of Russian Academy of Sciences, 620108 Ekaterinburg, Russia}

\begin{abstract}
Despite several decades of active experimental and theoretical studies of rare-earth orthoferrites, the mechanism of the formation of specific magnetic, magnetoelastic, optical, and magneto-optical properties remains a subject of discussion. 
This paper provides an overview of simple theoretical model approaches to quantitatively describing the structure-property relationships, in particular, interplay between FeO$_6$ octahedral deformations/rotations and main magnetic and optic characteristics such as N\'{e}el temperature, overt and hidden canting of magnetic sublattices, magnetic and magnetoelastic anisotropy, optic and photoelastic anisotropy.
\end{abstract}


\maketitle

\section{Introduction}

The rare-earth orthoferrites of the general formula
RFeO$_3$  (R = Y, or rare-earth ion) have attracted
 and continue to attract the particular attention of
researchers for several decades owing to their weak ferromagnetism, remarkable magneto-optical properties, spin-reorientation transitions
between antiferromagnetic phases, high velocity of domain walls, and many other properties. Their physical properties remain a
focus of considerable research due to promising applications in innovative spintronic devices, furthermore, they
contribute to an emerging class of materials, i.e., multiferroics with  strong magnetoelectric (ME) coupling.

The rare-earth orthoferrites RFeO$_3\,$ (space group D$_{2h}^{16}$) have a
distorted perovskite structure with only one type of Fe$^{3+}$ ion
octahedrally coordinated with six O$^{2-}$ ions \cite{Geller}.
Their complex non-collinear magnetic
structures and magnetic phase transitions are primarily due to the
combination of the antiferromagnetic (AFM) exchange
interaction with the Dzyaloshinskii-Moriya (DM) antisymmetric exchange interaction.

Despite the impressive advance of modern {\it ab-initio} band structure computation techniques, as
well as various quantum-chemistry cluster computation methods, for the majority of
practically important systems, these methods give only a rather rough picture of the
electronic structure and the energy spectrum which can at the best serve only as a
suggestion for building up an adequate theory. One of the most recent and illustrative examples of {\it ab-initio}  calculations is given in Ref.\,\cite{SFO}, the authors of which performed
 first principles simulations for the structural, elastic, vibrational, electronic and optical properties of orthorhombic samarium orthoferrite SmFeO$_3$ within the framework of density functional theory. However, such calculations encounter great difficulties in describing more "subtle" effects of magnetic, magnetoelastic,  and optical anisotropy\,\cite{CM-2019}.

Below, in the paper we apply  simple theoretical approaches to quantitatively evaluate the interplay between FeO$_6$ octahedral deformations/rotations and main magnetic and optic characteristics such as N\'{e}el temperature, overt and hidden canting of magnetic sublattices, magnetic and magnetoelastic anisotropy, optic and photoelastic anisotropy. These involve describing
a distorted (low-symmetry) structure as arising from
a (high-symmetry) parent structure with one or more
static symmetry-breaking structural distortions

Knowledge of the structure-property
relationships are required both to elucidate the nature of physical properties and to accelerate materials discoveries.

 Accessibility to computational methods make the distortion-mode analysis powerful, because it is possible to independently study
 various distortions  and directly assess their role
in forming the electronic structure and physical properties.

The structure–property relationships in ABO$_3$ perovskites exhibiting octahedral
rotations and  distortions were studied using group theoretical methods\,\cite{str-pro}.
Relationship between the rotation angles of octahedra
and bond-strength energy in crystals with perovskite structure was studied in Ref.\,\cite{Olekh}. However, there are practically no examples of using structure-property relationships to describe subtle effects of magnetic and optical anisotropy  in perovskites and, in particular, orthoferrites.

The immediate motivation to write this work was the recent publication of an article by Zhou {\it et al.}\,\cite{Zhou}, in which, contrary to seemingly well-established concepts, it is argued that single-ion anisotropy, and not the DM interaction, is responsible for the formation of the weak ferromagnetic moment of rare-earth orthoferrites. Moreover, the authors make this conclusion supposedly on the basis of the structure–property relationships. We will consistently show the incorrectness of their statements. 

The rest part of the paper is organized as follows. In Section 2 we address specific features of the crystal structure of rare-earth orthoferrites with a focus on the FeO$_6$ octahedra.  Sections 3 and 4 are devoted to intersite interactions, namely to Heisenberg superexchange interaction and Dzyaloshinskii-Moriya antisymmetric interaction which are mainly controlled by the FeO$_6$ octahedral rotations. In Sec.\,3 we shortly address main results of the microscopic theory of the isotropic superexchange interactions for  $S$-type Fe$^{3+}$ ions focusing on the angular dependence of the exchange integrals.
Most attention in Sec.\,4 centers around the derivation of the Dzyaloshinskii vector, its value, orientation, and sense (sign)  under different types of the (super)exchange interaction and crystal field. Theoretical predictions of this section are compared  with experimental data for the overt and hidden canting in orthoferrites.  Here, too, we consider a {\it weak ferrimagnetism}, a novel type of magnetic ordering in systems with competing signs of the Dzyaloshinskii vectors. 

In Secs. 5-6, we apply simple  theory to
quantitatively explore the relationship between structure
and different magnetic and optic properties of  rare-earth orthoferrites. 
In Sec.\,5 we address the "deformational" model of a single-ion magnetic and magnetoelastic anisotropy. In Sec.\,6 we develop the deformation model of linear birefringeance and  anisotropic photoelastic effects for orthoferrites. 
Short summary is presented in Sec.\,7.

\section{Crystal and magnetic structure of rare-earth orthoferrites}

Orthoferrites are composed of relatively robust
corner-shared FeO$_6$ octahedra, with nominally
12-coordinated rare earth (R) cations.  These adopt low-symmetry distortions from the ideal cubic perovskite Pm$\overline{3}$m symmetry.
The Fe atoms are located on inversion
centres, the R atom and atom O$_1$ lie on the mirror planes perpendicular to the orthorhombic $b$-axis, and
atom O$_2$ occupy a site of general symmetry.

The real FeO$_6$ complex in orthoferrites can be represented as a homogeneously
deformed ideal octahedron. To find the degree of distortion, we introduce a symmetric strain tensor according to the standard rules. In the local system of cubic axes of the octahedron
 \begin{equation}
\varepsilon_{ij}=\frac{1}{4l^2}\sum_{n=1}^6(R_i(n)u_j(n)+R_j(n)u_i(n)) \, ,
\label{u}
 \end{equation}
where ${\bf R}(n)$ is the radius-vector of the Fe-O$_n$ bond, ${\bf u}(n)$ is the O$_n$-ligand displacement vector, or
\begin{equation}
\hat{\varepsilon}=\left(
   \begin{array}{ccc}
    1-\frac{l_1}{l} & \frac{1}{2}(\frac{\pi}{2}-\theta_{12})&\frac{1}{2}(\frac{\pi}{2}-\theta_{13}) \\
	\frac{1}{2}(\frac{\pi}{2}-\theta_{21}) &1-\frac{l_2}{l} & \frac{1}{2}(\frac{\pi}{2}-\theta_{23}) \\
	\frac{1}{2}(\frac{\pi}{2}-\theta_{31})& 	\frac{1}{2}(\frac{\pi}{2}-\theta_{32}) &1-\frac{l_3}{l}	 
   \end{array}
  \right) \, ,
  \label{e}
 \end{equation}
 where $l$ is the Fe-O separation in an ideal octahedron, $l_i$ are the Fe-O$_i$ interatomic distances $\frac{1}{3}(l_1+l_2+l_3)=l$, and $\theta_{ij}$ are the bond angles O$_i$-Fe-O$_j$ in a real complex. Local $x,y,z$ axes in octahedron are defined as follows: the $z$-axis is directed along the Fe-O$_I$, the $x$-axis is along Fe-O$_{II}$ with the shortest Fe-O bond length. In general, the deformations of octahedra in orthoferrites are small and do not exceed 0.02.

Diagonal components of the traceless strain tensor (\ref{e}) (tensile/compressive deformations) can be termed as $E$-type deformations since $\varepsilon_{zz}$ and $\frac{1}{\sqrt{3}}(\varepsilon_{xx}-\varepsilon_{yy})$  transform according to the irreducible representation (irrep) $E$ of the cubic group O$_h$, while off-diagonal components (shear deformations) can be termed as $T_2$-type deformations since $var\epsilon_{yz}$,  $\varepsilon_{xz}$,  and  $\varepsilon_{xy}$  transform according to the irrep $T_2$ of the cubic group O$_h$.

The vector $\mathbf{v}$ of rotation of the octahedron FeO$_6$ , the direction and value of which specifies the axis and the angle of rotation, respectively, is related to the small displacements of oxygen ions as follows
 \begin{equation}
\mathbf{v}=\frac{1}{4l^2}\sum_{n=1}^6[{\bf R}(n)\times {\bf u}(n)] \, .
 \end{equation}

Four Fe$^{3+}$ ions occupy positions 4b in the orthorhombic elementary cell of orthoferrites RFeO$_3$ (space group $Pbnm$):
$$
1\,(1/2, 0, 0); 2\,(1/2, 0, 1/2); 3\,(0, 1/2, 1/2); 4\,(0, 1/2, 0)\,   .
$$
It is worth noting that another labeling
of the Fe$^{3+}$  positions than what was used here is found in the literature (see, for example, Refs.\,\cite{muon,Amelin}), in which case the basis vectors ${\bf G}$, ${\bf C}$, ${\bf A}$
may differ in sign.

Classical basis vectors of magnetic structure for 3$d$ sublattice are defined as follows:
$$
4S{\bf F}={\bf S}_1+{\bf S}_2+{\bf S}_3+{\bf S}_4\,; \\
$$
$$
4S{\bf G}={\bf S}_1-{\bf S}_2+{\bf S}_3-{\bf S}_4\,; \\
$$
$$
4S{\bf C}={\bf S}_1+{\bf S}_2-{\bf S}_3-{\bf S}_4\,; \\
$$
\begin{equation}
4S{\bf A}={\bf S}_1-{\bf S}_2-{\bf S}_3+{\bf S}_4\,,
\end{equation}
where ${\bf S}_i$ is classical spin vector for Fe-ion in $i$-th position, $S$\,=\,5/2 is the spin value. 
Here ${\bf G}$ describes the main antiferromagnetic component (N\'eel vector), ${\bf F}$ gives the weak ferromagnetic moment (overt canting),  the weak antiferromagnetic
components ${\bf C}$ and ${\bf A}$ describe a canting without  net magnetic moment (hidden canting). Allowed spin configurations for 3$d$-sublattice are denoted as $\Gamma_1\,(A_x, G_y, C_z)$,
$\Gamma_2\,(F_x, C_y, G_z)$, $\Gamma_4\,(G_x, A_y, F_z)$, where the components given in
parentheses are the only ones different from zero.
The phase diagram of orthoferrites  indicates all but SmFeO$_3$ adopt $\Gamma_4$ at room temperature.

Competition between magnetic anisotropy of the Fe-sublattice and R-Fe exchange interaction leads to the spin-reorientational transitions $\Gamma_4 \rightarrow \Gamma_2$ or $\Gamma_1$
 depending on R as temperature decreases\,\cite{KP}. In all cases, the $G$-component is prevalent: $G\gg F,C,A$.

\begin{figure}[t]
\begin{center}
\includegraphics[width=8cm,angle=0]{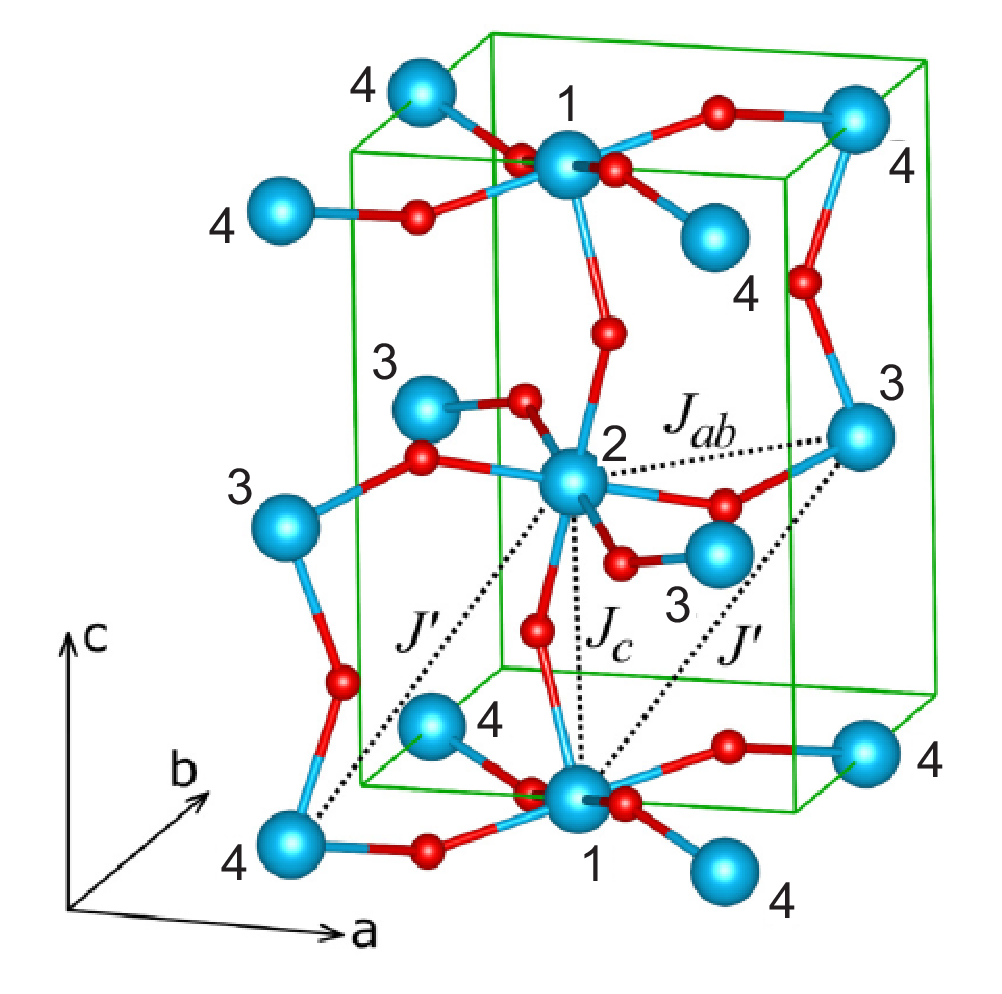}
\caption{(Color online) Structure of the Fe$^{3+}$--O$^{2-}$--Fe$^{3+}$ superexchange bonding in orthoferrites. $J_{ab}$ and $J_{c}$ are nearest-neighbor superexchange integrals, $J^{\prime}$ superexchange integral for next-nearest-neighbors. 1, 2, 3, 4, are Fe$^{3+}$ ions in four nonequivalent positions. Reproduced from Ref.\,\cite{Amelin}. }
\label{fig1}
\end{center}
\end{figure}

\section{Isotropic superexchange coupling and superexchange geometry}

Figure\,\ref{fig1} shows the intricate  structure
 of the Fe$^{3+}$--O$^{2-}$--Fe$^{3+}$ superexchange bondings in orthoferrites. 

\begin{figure}[t]
\centering
\includegraphics[width=8.5cm,angle=0]{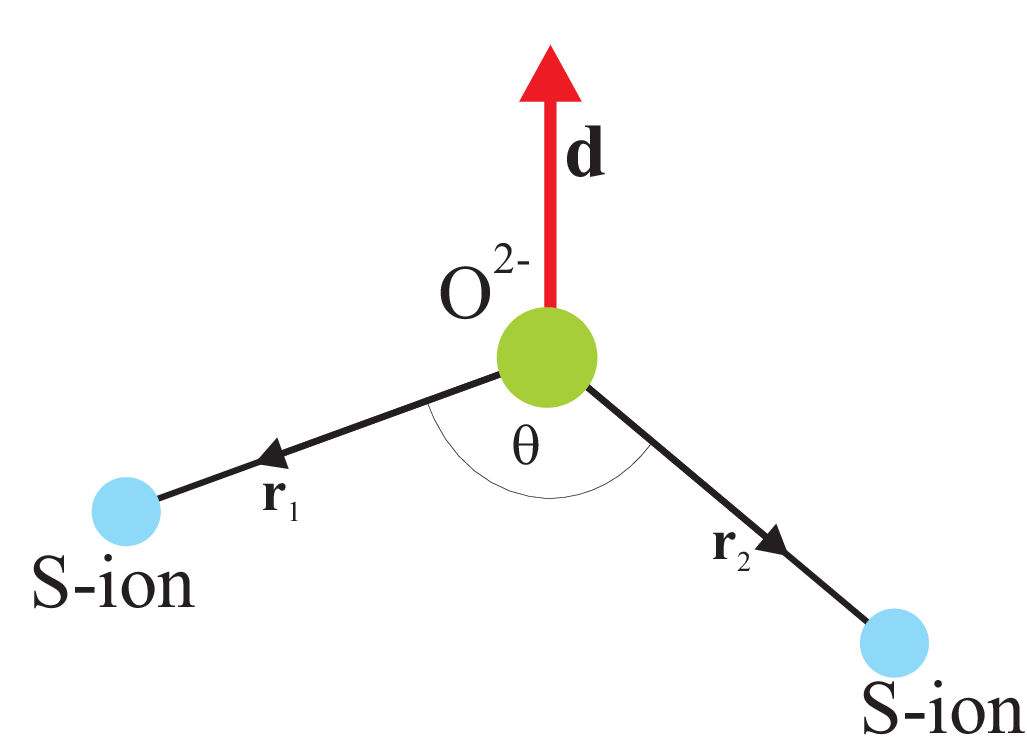}
\caption{Superexchange geometry and the Dzyaloshinskii vector.}
\label{fig2}
\end{figure}

The $G$-type of the magnetic structure of orthoferrites is determined by the strong isotropic superexchange interaction:
\begin{equation}
	{\hat V}_{ex}=\sum_{m>n}J_{mn}({\bf S}_m\cdot{\bf S}_n)	
\end{equation}
with an exchange integral, which depends primarily on the superexchange Fe-O-Fe bonding angle, which, in turn, is determined by the angles of the rigid rotation of the FeO$_6$ octahedra which produce deviations of the Fe-O-Fe bonding angle away from the ideal 180$^{\circ}$ found in the cubic perovskite aristotype (Pm$\overline{3}$m symmetry).

First poor man's microscopic derivation for the  dependence of the superexchange integral on the bonding angle (see Fig.\,\ref{fig1}) was performed by the author in 1970\,\cite{1971}  under simplified assumptions  for $S$-ions with configuration $3d^5$ (Fe$^{3+}$, Mn$^{2+}$)
\begin{equation}
	J_{12}(\theta )=a+b\cdot \cos\theta_{12} + c\cdot \cos^2\theta_{12} \, ,
	\label{angle}
	\end{equation}
where  parameters $a, b, c$ depend on the cation-ligand separation. A more comprehensive analysis  has supported validity of the expression. Interestingly, the second term in (\ref{angle}) is determined by the ligand inter-configurational $2p$-$ns$ excitations, while other terms are related with intra-configurational $2p$-, $2s$-contributions.

Later on the derivation had been generalized for the $3d$ ions in a strong cubic crystal field (see, e.g., Refs.\,\cite{Sidorov,Ovanesyan,thesis,JMMM-2016,CM-2019,JETP-2021})

Orbitally isotropic contribution to the superexchange integral for pair of Fe$^{3+}$, Cr$^{3+}$ ions with configurations $t_{2g}^{3}e_g^{2}$, $t_{2g}^{3}$, respectively,  can be written as follows
$$
J_{FeFe}=\frac{1}{25}(4J(e_ge_g)+12J(e_gt_{2g})+9J(t_{2g}t_{2g})) \,,
$$
$$
J_{CrCr}=J(t_{2g}t_{2g})\,,
$$
\begin{eqnarray}
J_{FeCr}=\frac{1}{5}(2J(e_ge_g)+3J(t_{2g}t_{2g}))\,.
\end{eqnarray}
Kinetic exchange contribution to partial exchange parameters $J(\gamma_i\gamma_j)$ related with the electron transfer to partially filled shells can be written as follows\,\cite{Sidorov,thesis}
$$
J(e_ge_g)=\frac{(t_{ss}+t_{\sigma\sigma}\cos\theta)^2}{2U};\,
J(e_gt_{2g})=\frac{t_{\sigma\pi}^2}{3U}\sin^2\theta;\,
$$
\begin{equation}
J(t_{2g}t_{2g})=\frac{2t_{\pi\pi}^2}{9U}(2-\sin^2\theta )\, ,
\label{kinetic}	
\end{equation}
where $t_{\sigma\sigma}>t_{\pi\sigma}>t_{\pi\pi}>t_{ss}$ are positive definite $d-d$ transfer integrals, $U$ is a mean $d-d$ transfer energy (correlation energy).
All the partial exchange integrals appear to be positive or "antiferromagnetic", irrespective of the bonding angle value,  though the combined effect of the $ss$ and $\sigma\sigma$ bonds $\propto \cos\theta$ in $J(e_ge_g)$ yields a ferromagnetic contribution given bonding angles $\pi /2<\theta <\pi$. It should be noted that the "large" ferromagnetic potential contribution\,\cite{Freeman} has a similar angular dependence\,\cite{Luk}.
\begin{figure}[t]
\centering
\includegraphics[width=8.5cm,angle=0]{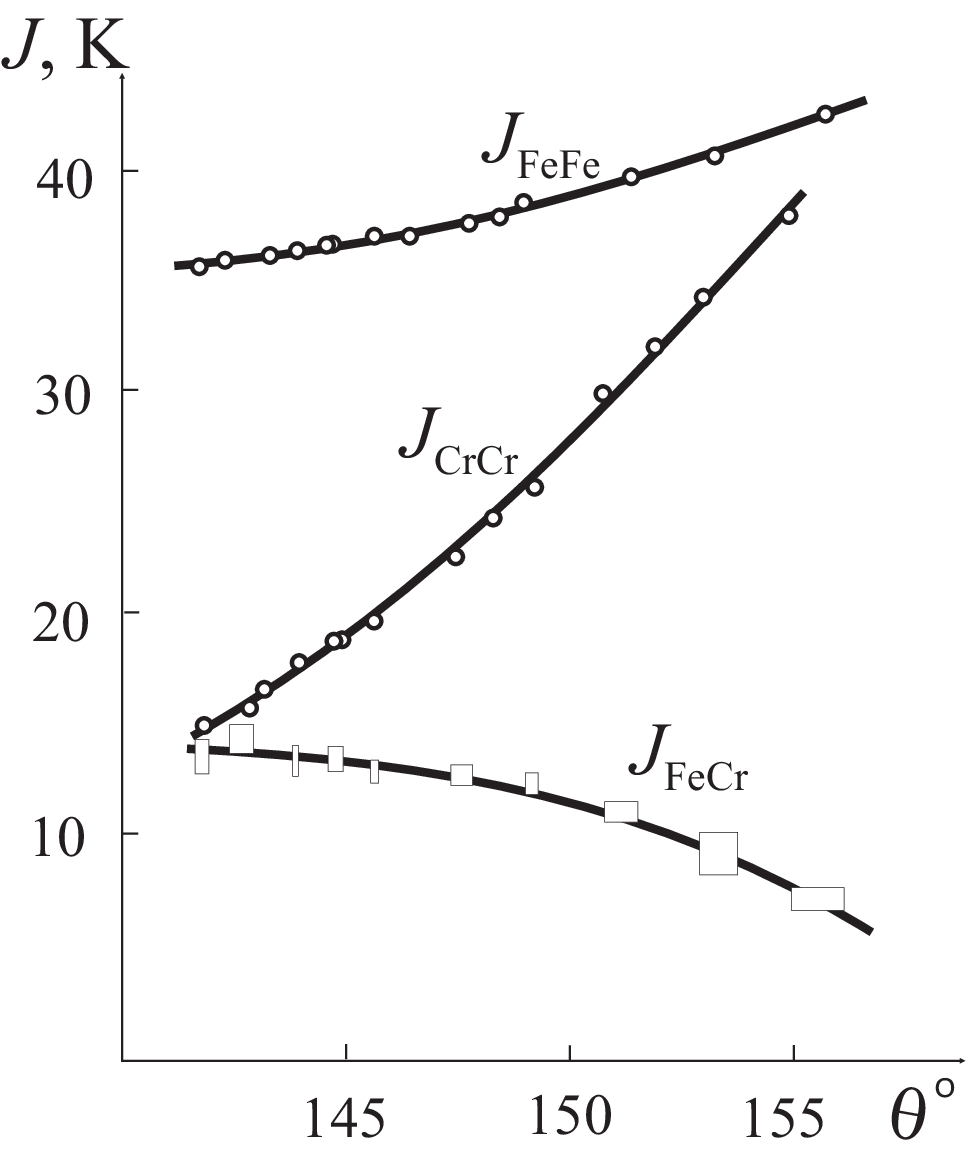}
\caption{Dependence of the Fe$^{3+}$-Fe$^{3+}$, Cr$^{3+}$-Cr$^{3+}$, Fe$^{3+}$-Cr$^{3+}$ exchange integrals (in K) on the superexchange bond angle in orthoferrites-orthocromites\,\cite{Ovanesyan}.}
\label{fig3}
\end{figure}

Some predictions regarding the relative magnitude of the $I(\gamma_i\gamma_j)$ exchange  parameters can be made using the relation among different $d-d$ transfer integrals as follows
\begin{equation}
	t_{\sigma\sigma}\,:\,t_{\pi\sigma}\,:\,t_{\pi\pi}\,:\,t_{ss}\approx \lambda_{\sigma}^2\,:\,\lambda_{\pi}\lambda_{\sigma}\,:\,\lambda_{\pi}^2\,:\,\lambda_{s}^2 \, ,
	\label{t-lambda}
\end{equation}
where $\lambda_{\sigma}, \lambda_{\pi}, \lambda_{s}$ are covalency parameters.
The simplified kinetic exchange contribution (\ref{kinetic}) related with the electron transfer to partially filled shells does not account for intra-center correlations which are of a particular importance for the  contribution related with the electron transfer to empty shells. For instance, appropriate contributions related with the transfer to empty $e_g$ subshell for the Cr$^{3+}$-Cr$^{3+}$ and Fe$^{3+}$-Cr$^{3+}$ exchange integrals are
$$
\Delta J_{CrCr}=-\frac{\Delta E(35)}{6U}\frac{t_{\sigma\pi}^2}{U}\sin^2\theta \, ;\,
$$
\begin{equation}
\Delta J_{FeCr}=-\frac{\Delta E(35)}{10U}\left[\frac{(t_{ss}+t_{\sigma\sigma}\cos\theta)^2}{U}+\frac{t_{\sigma\pi}^2}{U}\sin^2\theta\right] \, ,
\end{equation}
where $\Delta E(35)$ is the energy separation between $^3E_g$ and $^5E_g$ terms for $t_{2g}^3e_g$ configuration (Cr$^{2+}$ ion). Obviously, these contributions have a ferromagnetic sign. Furthermore, the exchange integral $J_{CrCr}$ can change sign  at $\theta$\,=\,$\theta_{cr}$:
\begin{equation}
	\sin^2\theta_{cr}=\frac{1}{\left(\frac{1}{2}+\frac{3}{8}\frac{\Delta E(35)}{U}\frac{t^2_{\sigma\pi}}{t^2_{\pi\pi}}\right)} \, .
\end{equation}

Microscopically derived angular dependence of the superexchange integrals does nicely describe the experimental data for exchange integrals $J_{FeFe}$, $J_{CrCr}$, and $J_{FeCr}$ in orthoferrites, orthochromites, and orthoferrites-orthochromites\,\cite{Ovanesyan} (see Fig.\,\ref{fig3}). The fitting allows us to predict the sign change for  $J_{CrCr}$ and $J_{FeCr}$ at $\theta_{12}$\,$\approx$\,133$^{\circ}$ and 170$^{\circ}$, respectively. In other words, the Cr$^{3+}$-O$^{2-}$-Cr$^{3+}$ (Fe$^{3+}$-O$^{2-}$-Cr$^{3+}$) superexchange coupling becomes ferromagnetic at $\theta_{12}\leq 133^{\circ}$ ($\theta_{12}\geq 170^{\circ}$).
However, it should be noted that too narrow (141-156$^{\circ}$) range of the superexchange bonding angles we used for the fitting with  assumption of the same Fe(Cr)-O bond separations and mean superexchange bonding angles for  all the systems gives rise to a sizeable parameter's uncertainty, in particular, for $J_{FeFe}$ and $J_{FeCr}$. In addition, it is necessary to note a large uncertainty regarding what is here called the "experimental"\, value of the exchange integral. The fact is that the "experimental"\, exchange integrals we have just used above are calculated using simple MFA relation for the N\'eel temperature
\begin{equation}
	T_N=\frac{zS(S+1)}{3k_B}J \, ,
\end{equation}
however, this relation yields the exchange integrals that can be one and a half or even twice less than the values obtained by other methods\,\cite{thesis,LuCrO3}.

Above we addressed only typically antiferromagnetic kinetic (super)exchange contribution as a result of the second order perturbation theory. However, actually this contribution does compete with typically ferromagnetic potential (super)exchange contribution, or Heisenberg exchange, which is a result of the first order perturbation theory. The most important contribution to the potential superexchange can be related with the intra-atomic ferromagnetic Hund exchange interaction of unpaired electrons on orthogonal ligand orbitals hybridized with the $3d$-orbitals of the two nearest magnetic cations.

Strong dependence of the $d-d$ superexchange integrals  on the cation-ligand-cation separation is usually described by the Bloch's rule\,\cite{Bloch}:
\begin{equation}
	\frac{\partial\ln J}{\partial\ln R}=\frac{\partial J}{\partial R}/\frac{J}{R} \approx -\,10 \, .
\end{equation}

\section{Crystal structure and the DM coupling in orthoferrites}

Weak ferromagnetism is one of the most remarkable physical properties of orthoferrites.
A theoretical explanation and first thermodynamic theory for weak ferromagnetism  was provided by I.~E. Dzyaloshinskii\,\cite{Dzyaloshinskii} in 1957 on the basis of symmetry considerations and Landau's theory of phase transitions of the second kind.

 Free energy of the two-sublattice uniaxial weak ferromagnet such as $\alpha$-Fe$_2$O$_3$, MnCO$_3$, CoCO$_3$, FeBO$_3$ was shown to be written as follows
\[
F=MH_E({\bf m}_1\cdot {\bf m}_2)-M{\bf H}_0({\bf m}_1+{\bf m}_2)+E_D+E_A
\]
\begin{equation}
	= MH_E({\bf m}^2-{\bf l}^2)-M{\bf H}_0{\bf m}+E_D+E_A \, .
\end{equation}
In this expression ${\bf m}_1$ and ${\bf m}_2$ are unit vectors in the directions of the sublattice moments, $M$ is the sublattice
magnetization, ${\bf m}=\frac{1}{2}({\bf m}_1+{\bf m}_2)$ and  ${\bf l}=\frac{1}{2}({\bf m}_1-{\bf m}_2)$ are the ferro- and antiferromagnetic vectors, respectively, $H_0$ is the applied field, $H_E$ is the exchange field,
$$
	E_D=-MH_D[{\bf m}_1\times {\bf m}_2]_z=+2MH_D[{\bf m}\times {\bf l}]_z= \\
$$
\begin{equation}
+2MH_D(m_xl_y-m_yl_x)
\end{equation}
is now called the Dzyaloshinskii interaction, $H_D$\,$>$\,0 is the Dzyaloshinskii field.
The anisotropy energy $E_A$ is assumed to have the form: $E_A = H_A/2M(m_{1z}^2+m_{2z}^2)=2H_A/2M(m_{z}^2+l_{z}^2)$, where $H_A$ is the anisotropy field. The choice of sign for the anisotropy field $H_A$ assumes that the $c$ axis is a hard direction of magnetization.
In a general sense the Dzyaloshinskii interaction implies the terms that are linear both on  ferro- and antiferromagnetic vectors. For instance, in orthorhombic orthoferrites and orthochromites the Dzyaloshinskii interaction consists of the antisymmetric and symmetric terms
$$
	E_D= d_1m_zl_x+d_2m_xl_z=
$$
$$
\frac{d_1-d_2}{2}(m_zl_x-m_xl_z)+\frac{d_1+d_2}{2}(m_zl_x+m_xl_z)=
$$
\begin{equation}
	-2MH_D[{\bf m}\times {\bf l}]_y+\frac{d_1+d_2}{2}(m_zl_x+m_xl_z)\, ,
\end{equation}
while for tetragonal fluorides NiF$_2$ and CoF$_2$ the Dzyaloshinskii interaction consists of the only symmetric term.
Despite Dzyaloshinskii supposed that weak ferromagnetism is due to relativistic spin-lattice and magnetic dipole interaction,  the theory was phenomenological one and did not clarify the microscopic nature of the Dzyaloshinskii interaction that does result in the canting.
 
 Later on, in 1960, T. Moriya\,\cite{Moriya} suggested a model microscopic theory of the exchange-relativistic antisymmetric exchange interaction to be a main contributing mechanism of weak ferromagnetism.
 He  extended the Anderson theory of superexchange to include spin-orbital coupling $V_{so}=\sum_i \xi({\bf l}_i\cdot{\bf s}_i)$, where $\xi$ is the coupling constant, and derived a spin-Hamiltonian  
\begin{equation}
V_{DM}=\sum_{mn}({\bf d}_{mn}\cdot\left[{\bf S}_m\times{\bf S}_n\right]) \, ,
	\label{DM}
\end{equation}
now called Dzyaloshinskii-Moriya (DM) spin coupling. Here,  ${\bf d}_{mn}$ is the axial
 Dzyaloshinskii vector.

Moriya found the symmetry constraints on the orientation of the Dzyaloshinskii vector $\mathbf{d}_{ij}$.
Let two ions 1 and 2 are located at the points A and B, respectively, with C point bisecting the AB line:

1. When C is a center of inversion: d=0.

2. When a mirror plane $\perp$AB passes through C,
     ${\bf d} \parallel$ mirror plane or  ${\bf d} \perp$ AB.

3. When there is a mirror plane including A and B,
     ${\bf d} \perp$ mirror plane.

4. When a twofold rotation axis $\perp$ AB passes through C,
    ${\bf d} \perp$ twofold axis.

5. When there is an n-fold axis (n$
\geq$2) along AB,  ${\bf d} \parallel$ AB.

Presently Keffer\,\cite{Keffer} proposed a simple phenomenological expression for the Dzyaloshinskii vector for two magnetic ions M$_i$ and M$_j$ interacting by the superexchange mechanism via intermediate ligand O (see Fig.\,\ref{fig1}):
\begin{equation}
	\mathbf{d}_{ij} \propto [\mathbf{r}_i \times \mathbf{r}_j] \, ,
\end{equation}
where ${\bf r}_{i,j}$ are unit radius vectors for O\,-\,M$_{i,j}$ bonds with presumably equal bond lenghts. Later on Moskvin\,\cite{1971} derived a microscopic formula for Dzyaloshinskii vector in a pair of the $S$-type ions
 \begin{equation}
	\mathbf{d}_{ij} = d_{ij}(\theta) [\mathbf{r}_i \times \mathbf{r}_j] \, ,
	\label{d12}
\end{equation}
where
\begin{equation}
	d_{ij}(\theta)=d_1(R_i,R_j)+d_2(R_i,R_j)\cos\theta_{ij}\, .
\end{equation}
In other words, at variance with superexchange integral, the Dzyaloshinskii vector depends both on the superexchange bonding angle and spatial Fe--O--Fe bond orientation. 


Note that the relation $d/J\approx \Delta g/g$, where $g$ is the gyromagnetic ratio, $\Delta g$ is its deviation from the free-electron value, proposed by Moriya for estimating the magnitude of the Dzyaloshinskii vector should be used with extreme caution. So, in the case of $S$-ions such as Fe$^{3+}$, Mn$^{2+}$, it is simply inapplicable.


 Spin nondiagonality of the DM coupling implies very unusual features of the ${\bf d}$-vector somewhat resembling vector orbital operator whose transformational properties cannot be isolated from the lattice. It seems  the ${\bf d}$-vector  does not transform as a vector at all.

Within simplest classical approximation the operator of symmetric and antisymmetric $d$\,-\,$d$ exchange interactions in orthoferrites
\begin{equation}
{\hat H}=\sum_{i>j}J_{ij}({\bf S}_i\cdot{\bf S}_j)+\sum_{i>j}({\bf d}_{ij}\cdot[{\bf S}_i\times{\bf S}_j])
\end{equation}
can be written in terms of basis vectors as a free energy  (see, e.g., Refs.\,\cite{thesis,Herrmann,JETP-2021} and references therein). Neglecting the terms quadratic in the components of small basis vectors ${\bf F}$, ${\bf C}$, ${\bf A}$, we obtain for the free energy per ion
$$
\Phi= \frac{J_F}{2}{\bf F}^2+\frac{J_G}{2}{\bf G}^2+\frac{J_C}{2}{\bf C}^2+\frac{J_A}{2}{\bf A}^2+
$$
\begin{equation}
D_x(C_yG_z-C_zG_y)+D_y(F_zG_x-F_xG_z)+D_z(A_xG_y-A_yG_x)
\,,
\label{EFGCA}
\end{equation}
where 
$$
J_F=-J_G=S^2(2J_{ab}+J_c); J_A=-J_C=S^2(2J_{ab}-J_c);
$$
$$
D_x=-S^2\sum_2d_x(12);
$$
$$
D_y=-S^2\left(\sum_4d_y(14)+\sum_2d_y(12)\right);
$$
 \begin{equation}
D_z=-S^2\sum_4d_z(14) \,,
\end{equation}
where $J_{ab}$ and $J_c$ are Fe-Fe exchange integrals in $ab$-plane and along $c$-axis, respectively (see Fig.\,\ref{fig1}), $d_{x,y,z}(ij)$ are the components of the Dzyaloshinskii vector for Fe$_i$-Fe$_j$ bond.  
By minimizing the free energy under condition ${\bf F}^2+{\bf G}^2+{\bf C}^2+{\bf A}^2=1$ and $F, C, A\ll G$ we find
$$
F_z=-\frac{D_y}{J_F-J_G}G_x;\,\, A_y=\frac{D_z}{J_A-J_G}G_x;
$$
$$
F_x=\frac{D_y}{J_F-J_G}G_z;\,\, C_y=-\frac{D_x}{J_C-J_G}G_z;
$$
\begin{equation}
A_x=-\frac{D_z}{J_A-J_G}G_y;\,\, C_z=\frac{D_x}{J_C-J_G}G_y;
\label{FGCA}
\end{equation}

Hereafter we address the DM coupling for the  S-type magnetic $3d$ ions with orbitally nondegenerate high-spin ground state in a strong cubic crystal field, that is for the $3d$ ions with half-filled shells $t_{2g}^3$, $t_{2g}^3e_g^2$, $t_{2g}^6e_g^2$ and ground states $^4A_{2g}$,  $^6A_{1g}$, $^3A_{2g}$, respectively.
The strong crystal field approximation seems to be more appropriate for the most part of $3d$ ions in crystals.

Making use of expressions for spin-orbital coupling $V_{so}$ and main kinetic contribution to the  superexchange parameters, that define the DM coupling,  after routine algebra we have found that the DM coupling can be written in a standard form (\ref{DM}) with Dzyaloshinskii vector (\ref{d12}), where  $d_{12}$ can   be written as follows\,\cite{1977,thesis,JMMM-2016,CM-2019,JETP-2021}
\begin{equation}
	d_{12}=X_1Y_2+X_2Y_1 \, ,
	\label{XY}
\end{equation}
where the exchange factors $X$ and dimensionless spin-orbital  factors $Y$ do reflect the exchange-relativistic structure of the second-order perturbation theory and details of the electron configurations for S-type ions.

The factors $X$ and $Y$ are presented in Table\,\ref{tableXY} for S-type 3d-ions. There  $\xi_{3d}$ is the spin-orbital parameter, $\Delta E_{^{2S+1}\Gamma}$ is the energy of the $^{2S+1}\Gamma$ crystal term, $t_{\sigma\sigma}>t_{\pi\sigma}>t_{\pi\pi}>t_{ss}$ are positive definite $d$\,-\,$d$ transfer integrals, $U$ is the $d$\,-\,$d$ transfer energy (correlation energy).

Note that the value of the Dzyaloshinskii vector in our approximation depends on the parameters of the FeO$_6$ octahedra rotation and does not depend at all on octahedral distortions, or on the parameters of the low-symmetry crystal field for Fe-centers, and hence on the $\delta g$ values, characterizing the deviation of the $g$- factor from its value in a free ion. Moriya's estimation $d/J\approx \Delta g/g$ in this case does not work at all. In particular, the authors of Ref.\,\cite{Zhou} incorrectly associate the value of $\delta{\hat g}=2\lambda{\hat \Lambda}$ (see expression (3) there) with the spin canting angle for orthoferrites.

The signs for $X$ and $Y$ factors in Table\,\ref{tableXY} are predicted for rather large superexchange bonding angles $|\cos\theta_{12}|>t_{ss}/t_{\sigma\sigma}$ which are typical for many $3d$ compounds such as oxides and a relation $\Delta E_{^4T_{1g}}(41)<\Delta E_{^4T_{1g}}(32)$ which is typical for high-spin $3d^5$ configurations.

On the whole, the data in Table\,\ref{tableXY} allow us to evaluate both the  numerical value and sign of the $d_{12}$ parameters.

It should be noted that for critical angle $\theta_{cr}$, when the Dzyaloshinskii vector changes its sign we have
$\cos\theta_{cr}=-d_1/d_2=\frac{\lambda_s^2}{\lambda_{\sigma}^2}$ for $d^8-d^8$ pairs and $	 \cos\theta_{cr}=-d_1/d_2=\frac{\lambda_s^2}{\lambda_{\sigma}^2-\lambda_{\pi}^2}$ for $d^5-d^5$ pairs. Making use of different experimental data for covalency parameters (see, e.g., Ref.\,\cite{Tofield}) we arrive at $d_1/d_2\sim \frac{1}{5}-\frac{1}{3}$ and $\theta_{cr}\approx 100^{\circ}-110^{\circ}$ for Fe$^{3+}$-Fe$^{3+}$ pairs in oxides.

 Relation among different $X$'s given the superexchange geometry and covalency parameters typical for orthoferrites and orthochromites \,\cite{thesis} is
\begin{equation}
	|X_{d^8}|\geq |X_{d^3}|\geq |X_{d^5}| \, ,
\end{equation}
however, it should be underlined its sensitivity both to superexchange geometry and covalency parameters. Simple comparison of the exchange parameters $X$ (see Table \ref{tableXY}) with exchange parameters $I(\gamma_i\gamma_j)$ (\ref{kinetic}) evidences their close magnitudes. Furthermore, the relation (\ref{t-lambda}) allows us to maintain more definite correspondence.

Theoretically predicted  signs of the Dzyaloshinskii vector in pairs of the S-type 3d-ions with local octahedral symmetry (the sign rules) are presented in Table\,\ref{tablesign}. The signs for $d^3-d^3$, $d^5-d^5$, and $d^3-d^8$ pairs turn out to be the same but opposite  to signs for $d^3-d^5$ and $d^8-d^8$ pairs.
In a similar way to how different signs of the conventional exchange integral determine different (ferro-antiferro) magnetic orders the different signs of the Dzyaloshinskii vectors  create a possibility of nonuniform  (ferro-antiferro) ordering of local weak (anti)ferromagnetic moments, or local overt/hidden cantings. Novel magnetic phenomenon and novel class of magnetic materials, which are systems such as solid solutions YFe$_{1-x}$Cr$_x$O$_3$ with competing signs of the Dzyaloshinskii vectors are discussed in Refs.\,\cite{JMMM-2016,JMMM-2018,CM-2019,JETP-2021}  in more detail.

\begin{widetext}
\begin{center}
\begin{table}
\caption{Expressions for the $X$ and $Y$ parameters that define the magnitude and the sign of the Dzyaloshinskii vector in pairs of the S-type 3d-ions with local octahedral symmetry. Signs for $X_i$ correspond to the bonding angle $\theta >\theta_{cr}$.}
\begin{tabular}{|c|c|c|c|c|c|}
\hline
   \begin{tabular}{c}
Ground state \\
configuration \\
\end{tabular}        & $X$ & Sign $X$ &  $Y$ & Sign $Y$ & \begin{tabular}{c}
Excited state \\
configuration \\
\end{tabular}  \\ \hline
  \begin{tabular}{c}
$3d^3$($t_{2g}^3$):${}^4A_{2g}$ \\
$V^{2+}$, $Cr^{3+}$, $Mn^{4+}$ \\
\end{tabular}  & $-\frac{1}{3U}t_{\pi\pi}t_{\sigma\pi}\cos\theta$ & + &$\frac{2\xi_{3d}}{3\sqrt{3}}(\frac{1}{\Delta E_{^4T_{2g}}}+\frac{2}{\Delta E_{^2T_{2g}}})$ & + & $t_{2g}^2e_g^1$ \\ \hline
 \begin{tabular}{c}
$3d^5$($t_{2g}^3e_g^2$):${}^6A_{1g}$ \\
$Mn^{2+}$, $Fe^{3+}$ \\
\end{tabular}  & \begin{tabular}{c}$-\frac{1}{5U}(t_{\pi\pi}t_{\sigma\pi}\cos\theta$ - \\
$t_{\pi\sigma}\left(t_{ss}+t_{\sigma\sigma}cos\theta )\right)$\\
\end{tabular} & -- & --$\frac{6\xi_{3d}}{5\sqrt{3}}(\frac{1}{\Delta E_{^4T_{1g}}(41)}-\frac{1}{\Delta E_{^4T_{1g}}(23)})$ & -- & $t_{2g}^4e_g^1$, $t_{2g}^2e_g^3$ \\ \hline	
\begin{tabular}{c}
$3d^8$($t_{2g}^6e_g^2$):${}^3A_{2g}$ \\
$Ni^{2+}$, $Cu^{3+}$ \\
\end{tabular}  & $\frac{1}{2U}t_{\pi\sigma}(t_{ss}+t_{\sigma\sigma}\cos\theta )$& -- & $\frac{3\xi_{3d}}{2\sqrt{3}}(\frac{1}{\Delta E_{^3T_{2g}}}+\frac{1}{\Delta E_{^1T_{2g}}})$ & + & $t_{2g}^5e_g^3$\\ \hline
	\end{tabular}
\label{tableXY}
\end{table}
\end{center}
\end{widetext}

\begin{center}
\begin{table}
\caption{Sign rules for the Dzyaloshinskii vector in pairs of the S-type 3d-ions with local octahedral symmetry and the bonding angle $\theta >\theta_{cr}$.}
\centering
\begin{tabular}{|c|c|c|c|}
\hline
   $3d^n$      & $3d^3$($t_{2g}^3$) &  $3d^5$($t_{2g}^3e_g^2$)  & $3d^8$($t_{2g}^6e_g^2$)  \\ \hline
  $3d^3$($t_{2g}^3$)  & + &--& + \\ \hline
 $3d^5$($t_{2g}^3e_g^2$) & -- &+ & + \\ \hline	
$3d^8$($t_{2g}^6e_g^2$) & +& +& -- \\ \hline
	\end{tabular}
\label{tablesign}
\end{table}
\end{center}

\begin{figure}[t]
\centering
\includegraphics[width=8.5cm,angle=0]{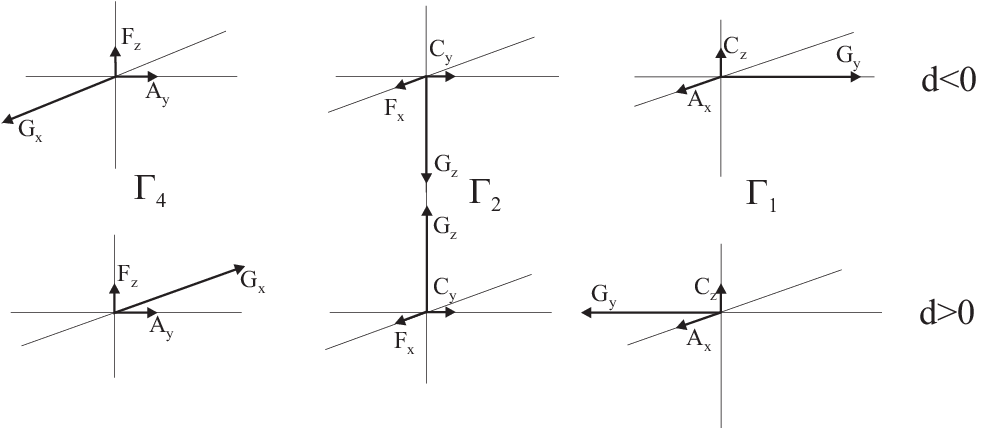}
\caption{Basic vectors of magnetic structure for $3d$ sublattice in orthoferrites and orthochromites}
\label{fig4}
\end{figure}

At variance with isotropic superexchange coupling the DM coupling has a much more complicated structural dependence. 
Figure\,\ref{fig3} shows the intricate  structure
 of the Fe$^{3+}$--O$^{2-}$--Fe$^{3+}$ superexchange bondings in orthoferrites that points to a complicated structural dependence of the Dzyaloshinskii vectors.

In Table\,\ref{tabler1r2}
we present structural factors $\left[{\bf r}_1\times{\bf r}_2\right]_{x,y,z}$ for the superexchange coupled Fe-O-Fe pairs in orthoferrites with numerical values for YFeO$_3$\,\cite{RFeO3}. In all cases, the vector ${\bf r}_1$ is oriented to the Fe ion in the position (1/2,0,0), the vectors ${\bf r}_2$ are oriented to the nearest Fe ions in the $ab$-plane (1a, 1b) or along the $c$-axis (3a). It is easy to see that the weak ferromagnetism in orthoferrites governed by the $y$-component of the Dzyaloshinskii vector does actually make use of only about one-third of its maximal value.
\begin{center}
\begin{table}
\caption{The structural factors $\left[{\bf r}_1\times{\bf r}_2\right]_{x,y,z}$ for the superexchange coupled Fe-O-Fe pairs in orthoferrites with numerical values for YFeO$_3$. See text for detail.}
\centering
\begin{tabular}{|c|c|c|c|}
\hline
         & $\left[{\bf r}_1\times{\bf r}_2\right]_x$ &  $\left[{\bf r}_1\times{\bf r}_2\right]_y$  & $\left[{\bf r}_1\times{\bf r}_2\right]_z$  \\ \hline
  1a  & $-\frac{z_2bc}{2l^2}$\,=\,-0.31 & $-\frac{z_2ac}{2l^2}$\,=\,-0.29 & $\frac{(y_2-x_2+\frac{1}{2})ab}{2l^2}$\,=\,0.41 \\ \hline
 1b & $+\frac{z_2bc}{2l^2}$\,=\,0.31 & $-\frac{z_2ac}{2l^2}$\,=\,-0.29 & $\frac{(y_2-x_2+\frac{1}{2})ab}{2l^2}$\,=\,0.41 \\ \hline	
3a & $\frac{(\frac{1}{2}-y_1)bc}{2l^2}$\,=\,0.20 &-$\frac{x_1ac}{2l^2}$\,=\,-0.55 & 0 \\ \hline
	\end{tabular}
\label{tabler1r2}
\end{table}
\end{center}
Simple formula for the Dzyaloshinskii vector (\ref{d12}) and structural factors from Table\,\ref{tabler1r2} can be used to find a relation between crystallographic and canted magnetic structures for four-sublattice's orthoferrites RFeO$_3$ and orthochromites RCrO$_3$\,\cite{1975,thesis,JMMM-2016,CM-2019,JETP-2021} (see Fig.\ref{fig3}), where main G-type antiferromagnetic order  is accompanied by both overt canting characterized by ferromagnetic vector ${\bf F}$ (weak ferromagnetism!) and two types of a hidden canting, ${\bf A}$ and ${\bf C}$ (weak antiferromagnetism!):
$$
F_z=\frac{(x_1+2z_2)ac}{6l^2}\frac{d}{J}G_x\,;\, F_x=-\frac{(x_1+2z_2)ac}{6l^2}\frac{d}{J}G_z\,;
$$
$$
A_y=\frac{(\frac{1}{2}+y_2-x_2)ab}{2l^2}\frac{d}{J}G_x\,;\,
A_x=-\frac{(\frac{1}{2}+y_2-x_2)ab}{2l^2}\frac{d}{J}G_y\,; 
$$
\begin{equation}
C_y=\frac{(\frac{1}{2}-y_1)bc}{2l^2}\frac{d}{J}G_z\,;\,C_z=-\frac{(\frac{1}{2}-y_1)bc}{2l^2}\frac{d}{J}G_y\,,
\label{FCA}	
\end{equation}
where $a,b,c$ are unit cell parameters, $x_{1,2}, y_{1,2}, z_2$ are oxygen ($O_{I,II}$) parameters\,\cite{RFeO3}, $l$ is a mean cation-anion separation. These relations imply an averaging on the Fe$^{3+}$-O$^{2-}$-Fe$^{3+}$ bonds in $ab$ plane and along $c$-axis. It is worth noting that $|A_{x,y}|>|F_{x,z}|>|C_{y,z}|$.

First of all we arrive at a simple relation between crystallographic parameters, canting angle, and  magnetic moment of the Fe-sublattice: in units of $G\cdot g/cm^3$
\begin{equation}
	M_{Fe}=\frac{4g_S\beta_eS}{\rho V}|F_{x,z}|=\frac{2g\beta_eSac}{3l^2\rho V}(x_1+2z_2)\frac{d(\theta)}{J(\theta)}\, ,
\end{equation}
where $\rho$ and $V$ are the unit cell density and volume, respectively.

\begin{figure}[t]
\centering
\includegraphics[width=8.5cm,angle=0]{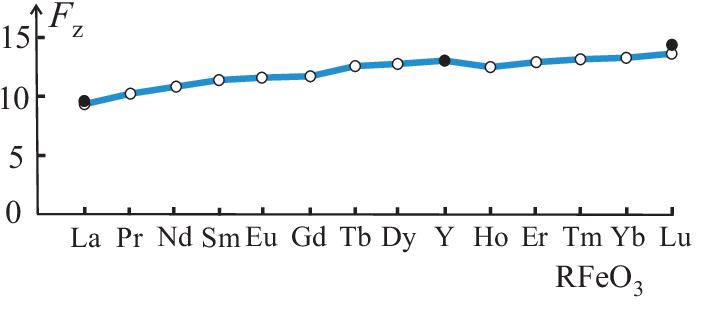}
\caption{Theoretical predictions for the overt canting in  orthoferrites (hollow circles) normalized on experimental data for YFeO$_3$. Solid circles are latest experimental data for orthoferrites with nonmagnetic R-ion\,\cite{Zhou}.}
\label{fig5}
\end{figure}

The theoretically predicted value of the spin canting angle along the $c$-axis, or $F_z$-component, increases monotonically from LaFeO$_3$ to LuFeO$_3$ (see Fig.\,\ref{fig5}), which is in excellent agreement with  latest experimental data obtained on precisely oriented crystals of orthoferrites with a nonmagnetic R-ion, R\,=\,La, Y, Lu\,\cite{Zhou}). The authors of Ref.\,\cite{Zhou} erroneously interpreted this dependence as evidence that "... single-ion anisotropy effect is responsible for the spin canting in the type-G antiferromagnets orthoferrites".

The overt canting $F_{x,z}$ can be calculated through the ratio of the Dzyaloshinskii ($H_D$) and exchange ($H_E$) fields as follows
\begin{equation}
	F=H_D/2H_E \, .
\end{equation}
If we know the Dzyaloshinskii field we can calculate the $d(\theta )$ parameter in orthoferrites as follows
\begin{equation}
	H_D=\frac{S}{g\mu_B}\sum_i|d_y(1i)|=\frac{S}{g\mu_B}(x_1+2z_2)\frac{ac}{l^2}|d(\theta )|\, ,
\end{equation}
that yields $|d(\theta )|\cong$\,3.2\,K in YFeO$_3$ given $H_D$\,=\,140\,kOe\,\cite{Jacobs}. It is worth noting that despite $F_z\approx$\,0.01 the $d(\theta )$ parameter is only one order of magnitude  smaller than the exchange integral in YFeO$_3$.

Our results have stimulated experimental studies of the  hidden canting, or "weak antiferromagnetism" in orthoferrites. As shown in Table\,\ref{AFC} the theoretically predicted relations between overt and hidden canting  nicely agree with the experimental data obtained for different orthoferrites by NMR\,\cite{Luetgemeier} and neutron diffraction \,\cite{Plakhtii,Georgieva}.

In all cases, the magnitude of the Dzyaloshinskii vector ${\bf d}_{12}$ is anticorrelated with the magnitude of the superexchange integral $J_{12}$ in the sense that the superexchange geometry, favorable for the former, is unfavorable for the latter.  The specific supersensitivity of the DM coupling to the superexchange geometry  allows us to consider this interaction, first of all, the value and orientation of the Dzyaloshinskii vector, as one of the most important indicators determining the role of structural factors.

\begin{widetext}
\begin{center}
\begin{table}
\caption{Hidden canting in orthoferrites.}
\centering
\begin{tabular}{|c|c|c|c|c|c|c|}
\hline
 Orthoferrite  & A$_y$/F$_z$, theory\,\cite{1975} & A$_y$/F$_z$, exp & A$_y$/C$_y$, theory\,\cite{1975} & A$_y$/C$_y$,  exp  \\ \hline
  YFeO$3$ &  1.10 & \begin{tabular}{c}
1.10\,$\pm$\,0.03\cite{Luetgemeier} \\
1.4\,$\pm$\,0.2\cite{Plakhtii}\\
1.1\,$\pm$\,0.1\cite{Georgieva}\\
\end{tabular} & 2.04 & ?\\ \hline
  HoFeO$3$ & 1.16
  & 0.85\,$\pm$\,0.10\cite{Georgieva} &  2.00&? \\ \hline
 TmFeO$3$ & 1.10
&  1.25\,$\pm$\,0.05\cite{Luetgemeier}& 1.83 & ?\\ \hline
 YbFeO$3$ & 1.11
&  1.22\,$\pm$\,0.05\cite{Plakhtii} & 1.79 & 2.0\,$\pm$\,0.2\cite{Luetgemeier}\\ \hline
 \end{tabular}
\label{AFC}
\end{table}
\end{center}
\end{widetext}
Determination of the "sign" of the Dzyloshinskii vector is of a fundamental importance from the standpoint of the microscopic theory of the DM coupling.
As was first shown in our paper\,\cite{sign} reliable local information on the sign of the Dzyaloshinskii vector, or to be exact, that of the scalar Dzyaloshinskii parameter $d_{12}$, can be extracted from the ligand NMR data in weak ferromagnets. The procedure was described in details for $^{19}$F NMR data in weak ferromagnet FeF$_3$\,\cite{sign}.
Theoretically simulated NMR spectrum does nicely agree with the experimental ones only for the "right" mutual orientations of the ${\bf F}$ and ${\bf G}$ vectors, that means $d(FeFe)>0$ in a full accordance with our theoretical sign predictions (see Table\,\ref{tablesign}).
The same result, $d(FeFe)>0$ follows from the the magnetic $x$-ray scattering amplitude measurements in the weak ferromagnet FeBO$_3$\,\cite{Dmitrienko}.

	\subsection{The DM coupling and effective magnetic anisotropy}
At variance with the {\it spin-symmetric} single-ion anisotropy and anisotropic exchange, the Dzyaloshinskii-Moriya interaction is a source of {\it spin-antisymmetric} anisotropy.
	Hereafter we demonstrate a contribution of the DM coupling into effective magnetic anisotropy in orthoferrites within a simple classical approach. Taking into account the expression (\ref{EFGCA}) for the classical energy of orthoferrite and relations (\ref{FGCA}) for small basis vectors	
	the classical energies of the three spin configurations  $\Gamma_1(A_x,G_y,C_z)$, $\Gamma_2(F_x,C_y,G_z)$, and $\Gamma_4(G_x,A_y,F_z)$  given  $|F_x|=|F_z|=F$, $|C_y|=|C_z|=C$, $|A_x|=|A_z|=A$ can be written as follows\,\cite{thesis,CM-2019,JETP-2021}	
	\begin{eqnarray}
E_{\Gamma_1}= J_{G}-48JS^2F^2\left[\frac{1}{3}(\frac{C}{F})^2+\frac{2}{3}(\frac{A}{F})^2\right] \, ; \\
E_{\Gamma_2}= J_{G}-48JS^2F^2\left[1+\frac{1}{3}(\frac{C}{F})^2\right] \, ; \\
E_{\Gamma_4}= J_{G}-48JS^2F^2\left[1+\frac{2}{3}(\frac{A}{F})^2\right] \, ,	
	\end{eqnarray}
	with obvious relation $E_{\Gamma_4} < E_{\Gamma_1}\leq E_{\Gamma_2}$.
	The energies allow us to find the constants of the in-plane magnetic anisotropy $E_{an}=k_1\,\cos2\theta$ ($ac$, $bc$ planes, $\theta$ is the polar angle of the ${\bf G}$-vector), $E_{an}=k_1\,\cos2\varphi$ ($ab$ plane, $\varphi$ is the azimutal angle of the ${\bf G}$-vector): $k_1(ac)=\frac{1}{2}(E_{\Gamma_2}-E_{\Gamma_4})$;  $k_1(bc)=\frac{1}{2}(E_{\Gamma_2}-E_{\Gamma_1})$;
$k_1(ab)=\frac{1}{2}(E_{\Gamma_4}-E_{\Gamma_1})$. Detailed analysis of different mechanisms of the magnetic anisotropy of the orthoferrites\,\cite{thesis,CM-2019,JETP-2021} points to a leading contribution of the DM coupling. Indeed, for all the orthoferrites RFeO$_3$ this mechanism does  predict a minimal energy for $\Gamma_4$ configuration which is actually realized as a ground state for all the orthoferrites, if one neglects the R-Fe interaction. Furthermore, predicted value of the constant  of the magnetic anisotropy in $ac$-plane for YFeO$_3$ $k_1(ac)$\,=2.0$\cdot 10^5$\,erg/cm$^3$ is close enough to experimental value of 2.5$\cdot 10^5$\,erg/cm$^3$\,\cite{Jacobs}.
Interestingly, the model predicts a close energy for $\Gamma_1$ and $\Gamma_2$ configurations so that $|k_1(bc)|$ is about one order of magnitude less than $|k_1(ac)|$ and  $|k_1(ab)|$ for most orthoferrites\,\cite{thesis,CM-2019,JETP-2021}. It means the anisotropy in $bc$-plane will be determined by a competition of the DM coupling with relatively weak contributors such as magneto-dipole interaction and single-ion anisotropy. It should be noted that the sign and value of the $k_1(bc)$ is of a great importance for the determination of the type of the domain walls for orthoferrites in their basic $\Gamma_4$ configuration (see, e.g., Ref.\,\cite{DyFeO3}).

In conclusion, we emphasize once again that we are considering the classical theory of the magnetic state of orthoferrites, which is the result of a simple MFA approximation. The applicability of this popular approximation to the description of quantum antiferromagnets with the Dzyaloshinskii interaction, in particular the anisotropy effects, raises natural doubts. For example, in the author's paper\,\cite{DM}, the role of the DM interaction as a source of magnetic anisotropy is considered in detail and it is shown that for quantum $s$\,=\,1/2 antiferromagnets in contrast to the simple MFA approach, the DM contribution to the energy of anisotropy for an exchange-coupled spin-1/2 pair turns into zero. However, just as in the case of isotropic exchange, the use of the classical description of the DM interaction for magnets with a large spin $S$\,=\,5/2 seems quite reasonable.

\section{Magnetic and magnetoelastic anisotropy in orthoferrites}

\subsection{Second-order spin anisotropy}
Free energy of the  second-order spin anisotropy of the Fe-sublattice in orthoferrites can be written as follows\,\cite{Herrmann,thesis}
$$
 \Phi_{an}^{(2)}=D G_z^2+E(G_x^2-G_y^2)+2p(G_xA_y+G_yA_x)+
$$ 
\begin{equation}
2q(G_yC_z+G_zC_y)+2r(G_xF_z+G_zF_x)	 \, ,
\end{equation}
where we confined ourselves to terms that are linear and quadratic in the components of the main antiferromagnetic vector $\bf G$.
First, let us pay attention to the appearance of three terms of the type of symmetric Dzyaloshinskii interaction, the inclusion of which leads to "symmetric" corrections in the expressions for the parameters of the overt and hidden canting (\ref{FGCA}) and (\ref{FCA}). In particular, taking into account the $r$-contribution leads to the appearance of a difference between the $F_x$ and $F_z$ weak-ferromagnetic components:
\begin{equation}
 \frac{F_z-F_x}{F}=\frac{4r}{D_y} .
\end{equation}
The main, quadratic in the components of the antiferromagnetic vector, contribution to the energy of magnetic anisotropy is usually considered by limiting the rotation of the vector $\bf G$ in a certain plane:
\begin{equation}
 \Phi_{an}^{(2)}=k_1\,\cos2\theta	
\end{equation}
for $ac$-, $bc$-planes or
\begin{equation}
 \Phi_{an}^{(2)}=k_1\,\cos2\varphi	
\end{equation}
for $ab$-plane.

The main mechanisms of second-order spin anisotropy for 3d-sublattice within a two-sublattice model are associated with single-ion anisotropy (SIA), as well as two-ion anisotropy, determined by the Dzyaloshinskii-Moriya coupling, magnetic dipole and  exchange-relativistic Fe-Fe interaction (TIA) (see, e.g., Refs.\,\cite{md,TIA}.)
Anisotropy parameters are not equal
in the two-sublattice and four-sublattice models because the
weak antiferromagnetic order is absorbed into renormalized
anisotropy parameters. Therefore anisotropy parameters
should not be directly compared between two and four
sublattice models. Two-sublattice model interpretation is typical for conventional magnetic "macroscopic"\, measurements, while spin waves excitations  measured by  the method of submillimeter dielectric (THz) spectroscopy\,\cite{Volkov,Amelin}, Raman scattering\,\cite{White-1982} or inelastic neutron scattering\,\cite{Hahn,Park}
 should been analyzed within a full four-sublattice model.

The second-order single-ion spin anisotropy for the $S$-type $3d$ ions is  a result of the third-order perturbation theory with a zero approximation corresponding to either a free ion or a highly symmetric cubic environment, taking into account the quadratic effects in the spin-orbit interaction and linear in the low-symmetry crystal field (LSCF).
\begin{equation}
 V_{SIA}=\sum_{\gamma\nu}d_{\gamma}	B_{\gamma\nu}^*{\hat V}^{2\gamma}_{\nu}(S) \, ,
\end{equation}
where ${\hat V}^{2\gamma}_{\nu}(S)=\sum_{q}\alpha^{\gamma\nu}_{q}{\hat V}^2_q$ are combinations  of the components of the rank-2 spin irreducible tensor operator which  are transformed according to the irreducible representation of the $O_h$ point symmetry group, $\gamma$\,=\,$E,T_2$, $B_{\gamma\nu}$ are the low-symmetry crystal field parameters, $d_{\gamma}\propto\frac{\lambda^2}{(\Delta E)^2}$ are dimensionless parameters.

The low-symmetry crystal field can be  represented as the sum of the local “deformation” contribution associated with low-symmetry distortions of the FeO$_6$ octahedron and the nonlocal contribution of the rest of the lattice.
Within the framework of the "deformation" model, the LSCF parameters
for $S$-type $3d$ ions in weakly distorted octahedra in the linear approximation are proportional to the components of the deformation tensor of the octahedron of the corresponding symmetry
\begin{equation}
 	B_{\gamma\nu}=b_{\gamma} \varepsilon^{\gamma}_{\nu} \, ,
\end{equation}
where $b_{\gamma} $ are the parameters of the electron-lattice coupling. Numerical estimates for $3d^5$ configuration\,\cite{Licht}  show that the $b_{E}$ parameters are about an order of magnitude higher than the $b_{T_2}$ parameters.
Thus, the Hamiltonian of the single ion spin anisotropy of the
second order can be represented as follows
\begin{equation}
 V_{SIA}=\sum_{\gamma\nu}K_{\gamma}	\left(\varepsilon^{\gamma}\cdot V^{2\gamma}(S)\right)\, ,
\end{equation}
where $K_{\gamma}=b_{\gamma}d_{\gamma}$. Within the mean-field approximation (MFA) for the energy of magnetic anisotropy we arrive at
 \begin{equation}
 E_{SIA}=\sum_{\gamma}\tilde{K}_{\gamma}	\left(\varepsilon^{\gamma}\cdot C^{2\gamma}({\bf S})\right)\, ,
\end{equation}
where $C^{2\gamma}({\bf S})$ is a symmetrized combination of tensorial spherical harmonics with classical vector ${\bf S}$ to be its argument,  $$
\tilde{K}_{\gamma} =K_{\gamma}\langle\langle V^2_0(S)\rangle\rangle=K_{\gamma}\frac{\langle\langle 3S_z^2-\frac{35}{4}\rangle\rangle
}{4\sqrt{3\cdot 5\cdot 7}}
$$
are in fact local temperature-dependent magnetoelastic constants.
The latter expression can be represented in a cartesian form as
$$
 E_{SIA}=\sqrt{\frac{3}{2}}\tilde{K}_{E}(\varepsilon_{11}\alpha_1^2+\varepsilon_{22}\alpha_2^2\,+
 \varepsilon_{33})\alpha_3^2)+
 $$
 \begin{equation}
 +\sqrt{6}\tilde{K}_{T_2}(\varepsilon_{23}\alpha_2\alpha_3+
 \varepsilon_{13}\alpha_1\alpha_3+
 \varepsilon_{12}\alpha_1\alpha_2)\, ,
 \label{E_SIA}
\end{equation}
where $\alpha_i$ are direction cosines of the vector ${\bf S}$ in the local system of cubic axes.
For weakly distorted octahedral Fe$^{3+}$O$_6$ complexes in orthoferrites RFeO$_3$ ($|\varepsilon_{ij}|\sim 10^{-2}$): $\tilde{K}_{E}\approx$\,20\,$cm^{-1}$, $\tilde{K}_{T_2}\approx$\,2.5\,$cm^{-1}$\,\cite{thesis,JETP-1981}.

In the system of crystallographic axes $a, b, c$,
we obtain an expression for the free energy of the single-ion crystallographic anisotropy as follows
\begin{equation}
 E_{SIA}=\sum_{\gamma}\tilde{K}_{\gamma}	\left(a^2(\gamma)\cdot C^2({\bf S})\right)\, ,
\end{equation}
where
\begin{equation}
 a^2_q(\gamma)=\sum_{\nu}D^{(2)}_{q;\gamma\nu}(\omega) \varepsilon^{\gamma}_{\nu}
\end{equation}
are structure factors which depend both on the FeO$_6$ octahedron rotation and deformation parameters, $D^{(2)}_{q;\gamma\nu}(\omega)=\sum_{q_1}\alpha^{\gamma\nu}_{q_1}D^{(2)}_{qq_1}(\omega)$ are linear combinations of Wigner matrices, $\omega=(\phi_1,\theta,\phi_2)$ are Euler angles, which determine transformation between octahedron local coordinates and $abc$-system.

\begin{center}
\begin{table}
\caption{Contributions of the main mechanisms to the first constants of the magnetic anisotropy of orthoferrites YFeO$_3$ and LuFeO$_3$ ($\times 10^5$\,erg/cm$^3$). See text for detail.}
\centering
\begin{tabular}{|c|c|c|c|c|c|c|}
\hline
 Mechanism  & \begin{tabular}{cc}
\multicolumn{2}{c}{$k_1(ac)$}\\
Y & Lu \\
\end{tabular}
& \begin{tabular}{cc}
\multicolumn{2}{c}{$k_1(bc)$}\\
Y & Lu \\
\end{tabular} & \begin{tabular}{cc}
\multicolumn{2}{c}{$k_1(ab)$}\\
Y & Lu \\
\end{tabular}  \\ \hline
 DM coupling & 3.1\quad 3.1 &-0.8\quad -0.9 & -3.9\quad -4.0  \\ \hline
  Magnetodipole &  0.9\quad 0.8 &-0.2\quad -0.5 & -1.1\quad -1.3 \\ \hline
 SIA & -1.9\quad 1.0 &-5.6\quad -1.8 & -3.7\quad -2.8  \\ \hline
 Total &  2.1\quad 4.9 &-6.6\quad -3.2 & -8.7\quad -8.1  \\ \hline
Experiment &  2.1\quad $\sim$\,6.0 &-5.7\quad ? & -7.8\quad ?  \\ \hline
 \end{tabular}
\label{Y-Lu}
\end{table}
\end{center}

Magnetodipole interaction in orthoferrites was considered in Refs.\,\cite{Bidaux,md}. First of all, it should be noted that due to the symmetry of the Fe sublattice of orthoferrites, the magnetodipole interaction does not contribute to the Dzyaloshinskii interaction. For all orthoferrites, the magnetodipole interaction stabilizes the $\Gamma_4$ configuration, and the contribution to the anisotropy constants for all planes decreases monotonically by a factor of about 40 on going from LuFeO$_3$ to LaFeO$_3$, reflecting a decrease in orthorhombic distortions. Magnetodipole contribution to $k_1(ac)$ for YFeO$_3$ reaches a value of the order of 40\% of its experimental value. 

Theoretical estimations\,\cite{JETP-1981,thesis} for  main
contributions  to the first constants of the magnetic anisotropy of orthoferrites YFeO$_3$ and LuFeO$_3$ are presented in Table\,\ref{Y-Lu}. The SIA contribution includes taking into account both the main local contribution calculated in the framework of the deformation model and a small nonlocal lattice contribution calculated in the point charges model.

We do not attach much importance to the exact coincidence of the predicted and experimental\,\cite{KP} values of the constant $k_1(ac)$ for YFeO$_3$. More important is the theoretical prediction of an unexpectedly strong increase in this constant for LuFeO$_3$. The SIA contribution to $k_1(ac)$ partially compensates for the large contribution of the DM interaction in YFeO$_3$,  whereas in LuFeO$_3$ they add up. 
This result is confirmed by experimental data on the measurement of the threshold field $H_{SR}$ of spin reorientation $\Gamma_4\rightarrow\Gamma_2$ in the orthoferrite Lu$_{0.5}$Y$_{0.5}$FeO$_3$, in which $H_{SR}$\,=\,15\,T as compared to $H_{SR}$\,=\,7.5\,T in YFeO$_3$\,\cite{JETP-1981}.
Thus, one can estimate $k_1(ac)$ in LuFeO$_3$ as about three times as much as $k_1(ac)$ in YFeO$_3$.

The value of the ratio $k_1(ab)/k_1(ac)\approx$\,3.7 was estimated from the experimental data of Raman spectroscopy in YFeO$_3$\,\cite{White-1982}.

Unfortunately, despite numerous, including fairly recent, studies of the magnetic anisotropy of orthoferrites, we do not have reliable experimental data on the magnitude of the contributions of various anisotropy mechanisms.

Competition of various contributions in the temperature dependence of AFMR (antiferromegnetic resonance) frequencies and anisotropy constants in YFeO$_3$ was addressed in Ref.\,\cite{Mukhin}. However, the authors neglected to take into account the hidden canting modes in the thermodynamic potential (see expression (1) in their article), which did not allow an adequate description of the DM contribution to the anisotropy.
Let us pay attention to recent works on the determination of the parameters of the spin Hamiltonian in YFeO$_3$ from measurements of the spin-wave spectrum by the inelastic neutron scattering\,\cite{Hahn,Park} and terahertz absorption spectroscopy\,\cite{Amelin}. However, these authors started with a simplified spin-Hamiltonian that took into account only Heisenberg exchange, DM interaction, and single-ion anisotropy. Obviously, disregarding the magnetic dipole and exchange-relativistic anisotropy, the "single-ion anisotropy" constants found by the authors are some effective quantities that are not directly related to SIA.

Concluding the subsection let us note that the contribution of single-ion crystallographic anisotropy to
the Dzyaloshinskii interaction in orthoferrites does not exceed  1\%\,\cite{thesis}.

\subsection{Magnetoelastic coupling}

The common nature of the magnetic and magnetoelastic anisotropy leads to the fact that we must require from microscopic theory a simultaneous explanation of the numerical values both for the  anisotropy constants and magnetoelastic constants.

In the general case, magnetoelastic energy is understood as the part of the crystal energy that describes the coupling of the magnetic (spin) subsystem of the crystal with the crystal lattice and depends both on the macroscopic deformation and latent displacements of the Bravais sublattices, and on the parameters of the magnetic (spin) order -- magnetization, antiferromagnetism vectors, and other basis vectors of the structure. Magnetoelastic interactions are manifested, for example, in a change in the size and shape of the sample upon a change in the magnetic state (magnetostriction), as well as in a change in the magnetic state upon deformation of the sample. The nature of magnetoelastic interactions is associated with the dependence of the parameters of exchange interactions, magnetic anisotropy on crystallographic parameters - interatomic distances and bond angles.

The main role in the magnetoelastic effects  is played by the terms of the  energy, which are quadratic in the components of the largest of the basis vectors, the antiferromagnetic vector $\bf G$: 
\begin{equation}
\Phi_{me}\,=\,\Lambda_{ijkl}^0\varepsilon_{ij}G_k\,G_l\,
+\,\Pi_{kl}^n(\Gamma_{\nu})\,u_n(\Gamma_{\nu})\,G_k\,G_l\>,\label{Pme}
\end{equation}
where $G_k$, $G_l$ are components of the antiferromagnetic vector, $\varepsilon_{ij}$ is the tensor of macroscopic deformations, $u_n(\Gamma_{\nu})$ are components of the symmetrized vectors of "hidden"\, displacements of the Bravais sublattices ("internal distortions"), that alone do not lead to macroscopic deformation of the crystal, $\Lambda_{ijkl}^0$, $\Pi_{kl}^n(\Gamma_{\nu})$ are the tensors of magnetoelastic constants. The elastic energy of the crystal has a standard form
$$
\Phi_e\,=\,\frac{1}{2}\,C_{ijkl}^0\varepsilon_{ij}\,\varepsilon_{kl}\,+\,
C_{ij}^n(\Gamma_{\nu})\,u_n(\Gamma_{\nu})\varepsilon_{ij}\,+
$$
\begin{equation}
\frac{1}{2}\,C^{nm}(\Gamma_{\nu})\,u_n(\Gamma_{\nu})\,u_m(\Gamma_{\nu})\,.
\label{Pe}
\end{equation}

Generally speaking, hidden displacements $u_n(\Gamma_{\nu})$ can be associated with deformations:
\begin{equation}
 u_n(\Gamma_{\nu})=A_{ij}^n(\Gamma_{\nu})\varepsilon_{ij}   \, ,
 \label{A}
\end{equation}
where $A_{ij}^n(\Gamma_{\nu})$ is the so-called "inner stress"\, tensor, so that as a result, it is possible to use the renormalized energies $\Phi_e$ and $\Phi_{me}$, where only the components of the strain tensor $\varepsilon_{ij}$ will appear.

The equilibrium values of macroscopic deformations and displacements of the sublattices are found by minimizing the elastic and magnetoelastic energies.

The magnetostriction effects are  usually described by a simplified expression for magnetoelastic energy in the form as follows
$$
\Phi_{me}=\lambda_0[G_z^2+\gamma (G_x^2-G_y^2)]Tr{\hat \varepsilon}+\lambda_1G_z^2(\varepsilon_{zz}-\frac{1}{3}Tr{\hat \varepsilon})
+
$$
$$
\lambda_2G_z^2(\varepsilon_{xx}-\varepsilon_{yy})+\lambda_3(G_x^2-G_y^2)(\varepsilon_{zz}-\frac{1}{3}Tr{\hat \varepsilon})+
$$
$$
\lambda_4(G_x^2-G_y^2)(\varepsilon_{xx}-\varepsilon_{yy})+
$$
\begin{equation}
\mu_1G_yG_z\varepsilon_{yz}+\mu_2G_xG_z\varepsilon_{xz}+\mu_3G_xG_e\varepsilon_{xy} \, ,
\label{lambda}
\end{equation}
where $Tr{\hat \varepsilon}=(\varepsilon_{xx}+\varepsilon_{yy}+\varepsilon_{zz})$.
For a spin-reorientation transition in a certain plane of the orthoferrite this energy can be represented as follows\,\cite{KP}
\begin{equation}
\Phi_{me}=(L_a\varepsilon_{aa}+L_b\varepsilon_{bb}+L_c\varepsilon_{cc})\cos2\theta+\frac{1}{2}\mu\varepsilon_{\alpha\beta}\sin2\theta\, ,
\end{equation}
($\theta \rightarrow \varphi$ for $ab$-plane), where $L_{a,b,c},\mu$ are magnetoelastic constants expressed in terms of parameters $\lambda$'s and $\mu$'s from (\ref{lambda}), $\varepsilon_{\alpha\beta}=\varepsilon_{ac}, \varepsilon_{ab}, \varepsilon_{bc}$ are shear deformations for the corresponding planes.

Table\,\ref{me} presents the results of theoretical estimates of the contribution of various mechanisms to the magnetostriction constants in YFeO$_3$,  performed within the framework of absolutely the same approximations and the same parameters that were used to calculate the anisotropy constants in Table\,\ref{Y-Lu}\,\cite{thesis}.

Surprisingly, the DM interaction, being the main source of effective magnetic anisotropy in orthoferrites, practically does not make any noticeable contribution to the magnetostriction constants.

The symmetry of the magnetodipole interaction leads to the relationship between the magnetoelastic parameters in (\ref{lambda})\,\cite{md}:
$$
\lambda_2=\lambda_3;\,\,\, \lambda_1+3\lambda_4=-\frac{3}{4}(\mu_1+\mu_2+\mu3) \, .
$$
In the limit of an ideal perovskite structure
$$
\lambda_2=\lambda_3=0;\,\, \lambda_1:\lambda_4:\mu_1:\mu_2:\mu_3 = (-9):2:8:8:(-12) \, ,
$$
so that for the magnetoelastic parameters $L_{a,b,c}$ we have
\[
L_a:L_b:L_c=
   \begin{cases}
    (-1):(-5):6\,, \,(ac-plane); \\
	(-5):(-1):6\,, \, (bc-plane); \\
	(-4):4:0\,, \, (ab-plane)\, .	 
   \end{cases}
 \]
The magnetodipole contribution to the magnetoelastic parameters, varying relatively weakly in the series of orthoferrites, makes a noticeable, although not determining, contribution to the magnetostriction constants.

For the nonlocal contribution of the lattice in the model of point charges, the magnetoelastic parameters $L_{a,b,c},\mu_i$ in the ideal perovskite limit satisfy the relations 
\begin{equation}
\begin{cases}
    L_a=-L_b\,, L_c=0\,,\mu_2=-8L_a\,, \,(ac-plane); \\
	  L_a=-L_b\,, L_c=0\,,\mu_1=+8L_a\,, \, (bc-plane); \\
L_a=-L_b\,, L_c=0\,,\mu_3=0\,, \, (ab-plane)\, ,	 
   \end{cases}
\label{L}
\end{equation}
which are quite satisfactory for LaFeO$_3$\,\cite{thesis}.
On the whole, this mechanism, like the magnetodipole one 
makes a noticeable, although not determining, contribution to the magnetostriction constants.

The deformation model of spin anisotropy considered above provides the simplest example of a microscopic mechanism for the formation of magnetoelastic energy. Indeed, considering macroscopic crystal deformations instead of octahedral deformations in the expression (\ref{E_SIA}) for the single-ion spin anisotropy energy, we arrive at the magnetoelastic energy
$$
 E_{me}=\Lambda_E(\varepsilon_{11}\alpha_1^2+\varepsilon_{22}\alpha_2^2+
 $$
 \begin{equation}
 +\varepsilon_{33})\alpha_3^2)+\Lambda_{T_2}(\varepsilon_{23}\alpha_2\alpha_3+
 \varepsilon_{13}\alpha_1\alpha_3+
 \varepsilon_{12}\alpha_1\alpha_2)\, ,
\end{equation}
where $\alpha_i$ are direction cosines of vector ${\bf S}$ in local system of cubic axes, parameters $\Lambda_E$\,=\,$\sqrt{\frac{3}{2}}\tilde{K}_{E}$ and  $\Lambda_{T_2}$\,=\,$\sqrt{6}\tilde{K}_{T_2}$ are magnetoelastic constants.

\begin{widetext}
\begin{center}
\begin{table}
\caption{Contributions of the main mechanisms to the magnetostriction constants for YFeO$_3$ .}
\centering
\begin{tabular}{|c|c|c|c|c|c|c|}
\hline
 Mechanism  & \begin{tabular}{cccc}
\multicolumn{4}{c}{$ac$-plane}\\
L$_a$ & L$_b$ & L$_c$ & $\mu_2$  \\
\end{tabular}
& \begin{tabular}{cccc}
\multicolumn{4}{c}{$bc$-plane}\\
L$_a$ & L$_b$ & L$_c$ & $\mu_1$  \\
\end{tabular}
&  \begin{tabular}{cccc}
\multicolumn{4}{c}{$ab$-plane}\\
L$_a$ & L$_b$ & L$_c$ & $\mu_3$  \\
\end{tabular}
  \\ \hline
 DM coupling & 0.0\quad 0.0\quad 0.1\quad 0.0 & 0.0\quad 0.0\quad 0.1\quad 0.0 & 0.1\quad 0.1\quad 0.0\quad 0.0 \\ \hline
  Magnetodipole & 0.0\quad -0.4\quad 0.4\quad -1.1 & -0.4\quad -0.1\quad 0.5\quad -1.1 & -0.3\quad 0.3\quad 0.0\quad 1.8 \\ \hline
 SIA - E term & -1.6\quad -1.3\quad 2.9\quad 4.1 & -1.0\quad -2.2\quad 3.2\quad 1.9 & 0.6\quad -0.9\quad 0.3\quad 12.3  \\ \hline
 SIA - T$_2$ term & -0.4\quad 0.2\quad 0.2\quad 1.7 & 0.1\quad -0.2\quad 0.1\quad 2.0 & 0.5\quad -0.5\quad 0.0\quad 0.6 \\ \hline
 SIA - lattice & 0.4\quad -0.2\quad -0.2\quad-2.1 & -0.1\quad 0.1\quad -0.0\quad -2.0 & -0.5\quad 0.4\quad 0.1\quad 0.0   \\ \hline
 Total & -1.6\quad -1.7\quad 3.3\quad 2.6 & -1.4\quad -2.4\quad 3.8\quad 0.8 & 0.3\quad -0.7\quad 0.4\quad 14.7 \\ \hline
Experiment &  -1.6\quad  -1.3\quad 2.9\quad 2.4 & ?\qquad ?\qquad ?\qquad ? & ?\qquad ?\qquad ?\qquad ? \\ \hline
 \end{tabular}
\label{me}
\end{table}
\end{center}
\end{widetext}

The results of calculating the magnetoelastic constants performed at the same values of $\tilde{K}_{E}\approx$\,20\,$cm^{-1}$, $\tilde{K}_{T_2}\approx$\,2.5\,$cm^{-1}$ as in the case of single-ion anisotropy, the deformation model of single-ion magnetoelastic coupling, primarily the $E$-contribution, can be the leading mechanism of magnetostriction for $3d$-system in orthoferrites. It is the $E$-contribution of the deformation model that determines the anomalously high value of the magnetoelastic parameter $\mu_3$, and hence the anomalously large values of the shear deformation $\varepsilon_{ab}$ upon spin-reorientation in the $ab$-plane. 

\begin{figure}[t]
\begin{center}
\includegraphics[width=8.5cm,angle=0]{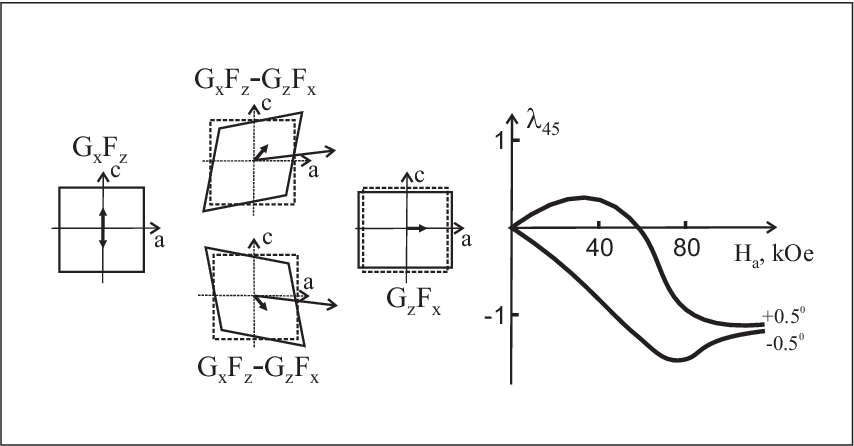}
\caption{(Color online) The field dependence of magnetostriction in YFeO$_3$ for the external field with orientation near ${\bf H}\parallel {\bf a}$\,\cite{JETPLett-1981}. Left in the figure - an illustration of the nature of shear deformation in antiferromagnetic domains with different orientations of the magnetic moment.}
\label{fig6}
\end{center}
\end{figure}

Theoretical predictions of the magnetoelastic "shear"\, parameters $\mu$ stimulated experimental studies of shear strains accompanying spin-reorientation transitions in orthoferrites\,\cite{JETPLett-1981}. A specific feature of such deformations is the dependence on the antiferromagnetic domain structure, so that in order to detect them during the $\Gamma_4-\Gamma_2$ transition induced by an external field, it was necessary to "violate" the exact orientation of the field along the $a$-axis of the crystal, thereby highlighting a certain type of domains (see Fig.\,\ref{fig6}).

A model quantitative analysis of the role of hidden displacements in magnetoelastic effects in orthoferrites was carried out in Ref.\,\cite{FTT-1987} based on calculations of elastic energy parameters within the framework of the rigid ion model.The authors showed that the contributions of the lattice strains and the contribution of the sublattice displacements to the magnetoelastic energy are comparable in magnitude, in agreement with the qualitative conclusions of Refs.\,\cite{Cullen,Bumagina}.

\subsection{Single-ion cubic anisotropy}

The fourth-order single ion spin anisotropy appears at least in the fourth order of the perturbation theory in the spin-orbit interaction and, in the general case, can be represented by an effective spin Hamiltonian as follows
\begin{equation}\label{Vsia4}
 	\hat{V}_{SIA}^{\mbox{\scriptsize \textit{(4)}}}
 	= \sum_{\gamma\nu}k_{\gamma\nu}^*{\hat V}^{4\gamma}_{\nu}(S) \, ,
\end{equation}
where we made use of cubic irreducible tensorial operators
${\hat V}^{4\gamma}_{\nu}(S)=\sum_{q}\alpha_{4q}^{\gamma\nu}{\hat V}^{4}_{q}(S)$, that is linear combinations of irreducible tensorial operators for rotation group acting in a spin space, which automatically "prohibits" the spin anisotropy of the fourth
order for $S$\,<\,2. In general, $\gamma =A_1,E,T_1,T_2$, however, for an ideal FeO$_6$ octahedron, the fourth-order spin anisotropy is actually reduced only to the cubic contribution  with $\gamma =A_1$, or cubic spin anisotropy:
$$
 	\hat{V}_{SIA}^{\mbox{\scriptsize \textit{cub}}} = k_{A_1}{\hat V}^{4A_1}_{0}(S)=
$$
 \begin{equation}
  =	k_{A_1}\left[\sqrt{\frac{7}{12}}{\hat V}^{4}_{0}(S)+\sqrt{\frac{5}{24}}({\hat V}^{4}_{4}(S)+{\hat V}^{4}_{-4}(S))\right]\,,
  \label{ka0}
\end{equation} 	
or in cartesian coordinates	
\begin{equation} 	
 \hat{V}_{SIA}^{\mbox{\scriptsize \textit{cub}}}	= \frac{a}{6}
 	\left[{\hat S}_{x}^{4} + {\hat S}_{y}^{4} + {\hat S}_{z}^{4}
 		- \frac{1}{5}\, S\left(S+1\right) \left(3S^{2}+3S-1\right)\right]\,	,
 \label{kA1}
\end{equation}
where $a=\,\frac{5\sqrt{3}}{12}k_{A_1}$ given $S$=5/2. 
In the mean-field approximation we obtain for the energy of magnetic cubic anisotropy 
\begin{equation}
 	E_{SIA}^{\mbox{\scriptsize \textit{cub}}} =  \tilde{k}_{A_{1}}C^{4A_1}_{0}({\bf S})  \,,
 	\label{kA2}
\end{equation}
where 
$$
C^{4A_1}_{0}({\bf S})=\sqrt{\frac{7}{12}}C^{4}_{0}({\bf S})+\sqrt{\frac{5}{24}}(C^{4}_{4}({\bf S})+C^{4}_{-4}({\bf S}))
$$
is the invariant cubic tensor spherical harmonic,
\begin{widetext}
\begin{equation}
  \tilde{k}_{A_{1}}=  k_{A_1}\langle\langle{\hat V}^{4}_{0}(S)\rangle\rangle =
k_{A_1}\frac{\langle\langle 35S_z^4-(30S^2+30S-25)S_z^2 +3S^4+6S^3-3S^2\rangle\rangle}{2\sqrt{(2S+5)(2S+3)(2S+1)(2S-1)(2S-3)(S+2)(S+1)S(S-1)}}  	\label{kA3}
\end{equation}
\end{widetext}

is the temperature-dependent anisotropy constant (S\,=\,5/2). 
Cubic spin anisotropy has
simple form (\ref{ka0}) or (\ref{kA1}), (\ref{kA2}) only in the coordinate system, where the  $xyz$-axes coincide with the principal axes of the cubic crystal field, that is symmetry axes of the fourth order. In the system of $abc$-axes, the energy of cubic anisotropy  has more complex expression
\begin{equation}
 	E_{SIA}^{\mbox{\scriptsize \textit{cub}}} = \tilde{k}_{A_{1}}\sum_{\Gamma\mu} a_{\Gamma\mu}C^{4\Gamma}_{\mu}({\bf S})  \,,
 	\label{kA4}
\end{equation}
where 
$$
C^{4\Gamma}_{\mu}({\bf S})=\sum_q\alpha_{4q}^{\Gamma\mu}C^{4}_{q}({\bf S})
$$
is the combination of spherical tensor harmonics to be a basis of the irrep $\Gamma$,
\begin{equation}
 	 a_{\Gamma\mu}=\sum_{qq'}\alpha_{4q}^{\Gamma\mu}{}^*D^{(4)}_{qq'}(\omega) \alpha_{4q'}^{A_10}  \,,
 	\label{kA5}
\end{equation}
are structure factors which depend on the FeO$_6$ octahedron rotation parameters, $D^{(4)}_{qq'}(\omega)$ are Wigner matrices, $\omega=(\phi_1,\theta,\phi_2)$ are Euler angles, which determine transformation between $abc$- and octahedron systems, that is octahedron rotation parameters.  

It is practically important to consider the cubic  spin
anisotropy for different crystal planes by replacing ${\bf S}\rightarrow {\bf G}$ in argument of spherical harmonic in (\ref{kA4}),  limiting the rotation of the vector $\bf G$ in a certain plane, and highlighting the fourth-order contribution:
\begin{equation}
 \Phi_{an}^{(4)}=k_2\,\cos4\theta	
\end{equation}
for $ac$-, $bc$-planes or
\begin{equation}
 \Phi_{an}^{(4)}=k_2\,\cos4\varphi	
\end{equation}
for $ab$-plane.

\begin{figure}[t]
\centering
\includegraphics[width=8cm,angle=0]{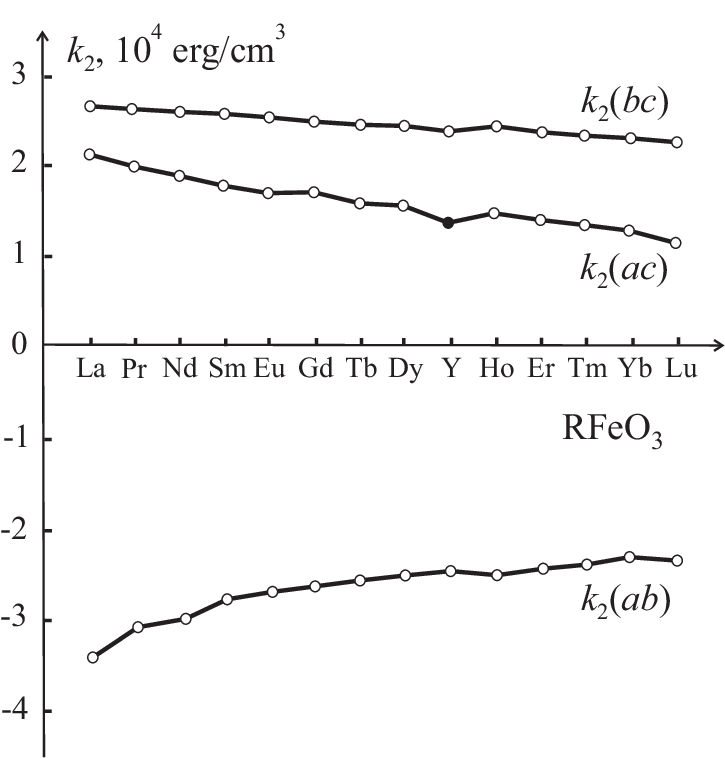}
\caption{The YFeO$_3$-normalized ($ac$-plane) $k_2$ constants for orthoferrites}
\label{fig7}
\end{figure}

Figure\,\ref{fig7} shows the calculated values of the fourth-order anisotropy constants for orthoferrites\,\cite{cubic} with the parameter $\tilde{k}_{A_{1}}$ normalized to the experimental value $k_2(ac)$ for YFeO$_3$: $k_2(ac)$\,=\,1.35$\cdot 10^4 erg/cm^3$\,\cite{KP}. 
On the whole, the constants $k_2$ rather smoothly decrease in absolute value (Fig.\,\ref{fig7}), changing by no more than two times on going from La to Lu. The difference between the constants $k_2(ac)$ and $k_2(bc)$ can serve as a measure of the deviation from the ideal cubic perovskite structure, for which $k_2(ac)=k_2(bc)=-\frac{3}{4}k_2(ab)$.
The different signs of these constants, positive for the $ac$-, $bc$-planes and negative for the $ab$-plane, indicate a different character of spin-reorientation transitions in the corresponding planes, that is second-order transitions in the $ac$-, $bc$-planes and first-order transitions in the $ab$-plane\,\cite{KP}. 
Indeed, all currently known spin-reorientation transitions of the $\Gamma_4 - \Gamma_2$ ($G_x - G_z$) type
in orthoferrites RFeO$_3$ (R = Sm, Nd, Er, Tm)
are smooth with two characteristic temperatures of the second order phase transitions to be a start and finish of the spin-reorientation, and the only known for these crystals
a transition of the type $\Gamma_4 - \Gamma_1$ ($G_x - G_y$) (DyFeO$_3$) is a jump-like transition
of the first type.
A unique example that confirms our conclusions about the sign of the second anisotropy constant is a mixed orthoferrite Ho$_{0.5}$Dy$_{0.5}$FeO$_3$\,\cite{KP}
in which two spin-reorientation
transitions $G_x - G_y$ ($T$\,=\,46\,K) and $G_y - G_z$ (18\,$\div$\,24\,K), realized through one phase transition of the first order in $ab$-plane and two phase transitions of the second order in $bc$-plane, respectively.

The microscopic expression for the cubic anisotropy constant $k_{A_1}$ for the ground state of ions Mn$^{2+}$ or Fe$^{3+}$ with $3d^5$ configuration, obtained in the scheme of a strong crystal field, looks as follows\,\cite{cubic}:
$$
k_{A_1}=\sum_{iS\Gamma}	 \left(\sum_j\frac{\lambda^{(1T_1)}_{{}^6A_{1g};j{}^4T_{1g}}\lambda^{(1T_1)}_{j{}^4T_{1g};i{}^{2S+1}\Gamma}}{\Delta E(j{}^4T_{1g})}\left\{
 	{\arraycolsep=0.2em
 	\begin{array}{ccc}
 		{1} & {1} & {2}
 		\\
 		{S} & {\frac{5}{2}} & {\frac{3}{2}}
 	\end{array}
 	} 	
 	\right\}\right)^2\times
 $$
 \begin{equation}
 \times (-1)^{\Gamma}\Delta E^{-1}(i{}^{2S+1}\Gamma )\left\{
 	{\arraycolsep=0.2em
 	\begin{array}{ccc}
 		{2} & {2} & {4}
 		\\
 		{\frac{5}{2}} & {\frac{5}{2}} & {S}
 	\end{array}
 	} 	
 	\right\}\,,
\end{equation}
where $i,j$ distinguish cubic terms, the ${}^{2S+1}\Gamma $ are excited terms ${}^4E_g(\times 2)$, ${}^2E_g(\times 7)$, ${}^2T_{2g}(\times 10)$, ${}^4T_{2g}(\times 3)$ (the number of identical terms is indicated in brackets), $E(i{}^{2S+1}\Gamma )$ is the term energy  measured from the energy of the ground ${}^6A_{1g}$ term, $\lambda^{(1T_1)}_{{}^6A_{1g};j{}^4T_{1g}},\,\lambda^{(1T_1)}_{j{}^4T_{1g};i{}^{2S+1}\Gamma}$ are spin-orbital parameters, $\left\{
 	{\arraycolsep=0.2em
 	\begin{array}{ccc}
 		{\cdot} & {\cdot} & {\cdot}
 		\\
 		{\cdot} & {\cdot} & {\cdot}
 	\end{array}
 	} 	
 	\right\}$ are 6$j$-symbols.
The main difficulty in calculating $k_{A_1}$  is associated
with a very large number of terms in the sums over $i, j, S, \Gamma $
as well as the complexity of calculating the reduced matrix
elements of the spin orbit taking into account the mixing of terms of the same symmetry.
Numerical calculations performed in the strong cubic field scheme for the Fe$^{3+}$ ion with
crystal-field parameter 10D$q$\,=\,12\,200\,$cm^{-1}$, Racah parameters B\,=\,700\,$cm^{-1}$, C\,=\,2600\,$cm^{-1}$, which correspond the Fe$^{3+}$ ion in orthoferrite YFeO$_3$\,\cite{Kahn,thesis}, yield
$$
k_{A_1}=(0.678 \zeta^4_{\pi\sigma}	+0.091 \zeta^3_{\pi\sigma}\zeta_{\pi\pi}-0.460 \zeta^2_{\pi\sigma}\zeta^2_{\pi\pi}-
$$
\begin{equation}
-\,0.045 \zeta_{\pi\sigma}\zeta^3_{\pi\pi}	-0.002 \zeta^4_{\pi\pi})\cdot 10^{-13} \,\,cm^{-1}\,,
\end{equation}
given $\zeta_{\pi\sigma}=-3\sqrt{2}\zeta_{3d}$, $\zeta_{\pi\pi}=3\zeta_{3d}$ and for the spin-orbital coupling constant  $\zeta_{3d}=500\,cm^{-1}$ yields
$$
k_{A_1}= 0.78 \,\, cm^{-1}\,.
$$
Introducing a single reduction factor for the parameters of the spin-orbital coupling $\kappa$\,=\,0.86, we obtain $k_{A_1}^*\approx$\,0.43\,$cm^{-1}$, which nicely agrees with the value calculated from experimental data for $k_2(ac)$ in YFeO$_3$ and other orthoferrites\,\cite{KP}.

\section{Optical anisotropy and anisotropic photoelastic effects in orthoferrites}

Hereafter we will show that simple models of the structure-property relationships, which were well proved above in the analysis of magnetic and magnetoelastic anisotropy, can be successfully used to analyze the optical and photoelastic anisotropy of orthoferrites.

\subsection{Natural birefringence of orthoferrites}

The analysis of the absorption spectra\,\cite{Kahn}, optical and magnetooptical anisotropy\,\cite{Z1,Z2} of orthoferrites  in a wide spectral range
 shows strong evidence for the key role of the dipole-allowed charge transfer (CT) $p-d$ transitions ${}^6A_{1g}\rightarrow{}^6T_{1u}$ in  the slightly distorted octahedral complexes FeO$_6$.


Optically, the orthoferrites are biaxial crystals showing a relatively
large natural birefringence\,\cite{Clover}. The comparative analysis of
the numerical values and the  frequency characteristics of birefringence for rare-earth
orthoferrites shows that the large natural birefringence
in orthoferrites at T\,=\,300\,K is mainly due to the $\,3d\,$-sublattice\,\cite{Tabor}.
In particular, the wavelength dependence of the $\,ab\,$-plane birefringence
($\Delta n_{ab}\,=\,n_a\,-\,n_b$) is basically {\it the same} in all orthoferrites
including YFeO$_3$\,\cite{Tabor}. Optical axes in Eu,\,Tb,\,Dy,\,Yb\,
orthoferrites, and the Y orthoferrite are inclined in fact at the same
angle of $\,\pm\>50\,^{\circ}\,$ to the $\,c\,$-axis ($\lambda \,=\,0.68\,\mu m$)\,\cite{Tabor,Chetkindi}.

\begin{figure}[htbp]
	\centering
		\includegraphics[width=8cm,angle=0]{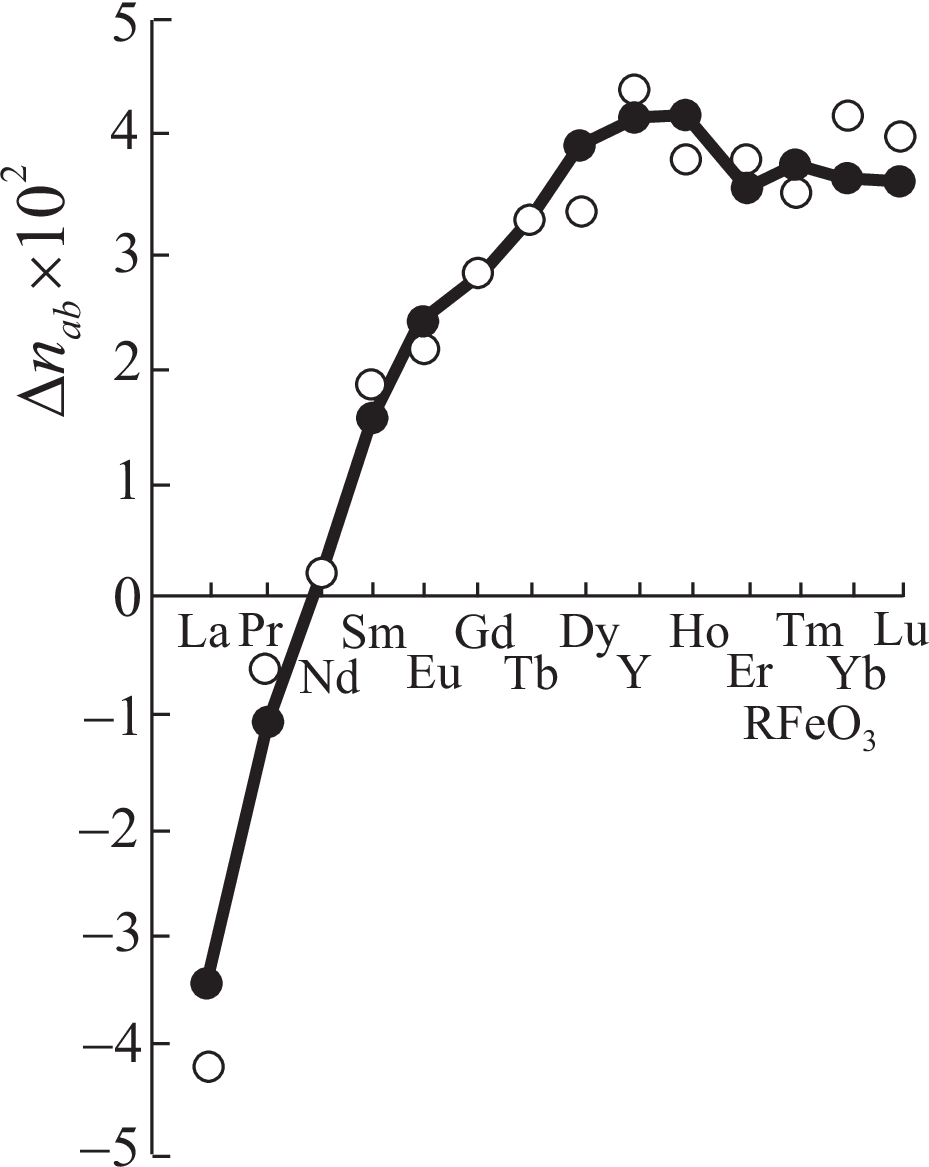}
		\caption{\footnotesize Linear birefringeance $\Delta n_{ab}$ for orthoferrites RFeO$_3$ in $ab$-plane: 
		solid circles are predictions of the deformation model, hollow circles are experimental data\,\cite{Clover}}
	\label{fig8}
\end{figure}

However, the natural birefringence in the $\,ab\,$-plane of orthoferrites at T\,$\approx$\,300\,K exhibits a puzzling behavior with a change in sign when passing from LaFeO$_3$ to LuFeO$_3$ with a more or less regular change in the value from -4$\cdot 10^{-2}$ to +4$\cdot 10^{-2}$\,\cite{Clover,Tabor} (see Fig.\,\ref{fig8}). Such a behavior can be related with 
 the specific behavior of distortions of the FeO$_6$ octahedra in the series of orthoferrites\,\cite{Clover1}. 
Indeed, the linear birefringence is determined by the anisotropic part of the permittivity tensor, which, in turn, for the contribution of $p-d$ CT transitions is determined by the anisotropic part of the FeO$_6$ octahedron polarizability tensor.
Within the "deformation model"\, the anisotropic  symmetric part of the polarizability tensor for the FeO$_6$ octahedron can be written as follows: 
\begin{equation}
\alpha_{ij}\,=\, \left\{
\begin{array}{cc}
p_E\,    \varepsilon_{ij}\,,&\>i\,=\,j \\
p_{T_2}\,\varepsilon_{ij}\,,&\>i\, \not =\,j\>,
\end{array} \right. \label{eq:65}
\end{equation}
where $\,\varepsilon_{ij}\,$ is the FeO$_6$-octahedron deformation tensor
($Tr\,\hat \varepsilon \,=\,$0); $\>p_{E,\,T_2}\,$ are the photoelastic constants, relating
the polarizability to $E\,,T_{2}$ -deformations, respectively. The relation
(\ref{eq:65}) is
valid in the local coordinate system of the FeO$_6$-octahedron. 
In the $\,abc\,$-axes system, it can be rewritten as
\begin{equation}
\alpha_{ij}\>=\>p_E\, \varepsilon_{ij}^E\>+\>\,p_{T_2} \varepsilon_{ij}^{T_2}\>,\label{eq:66}
\end{equation}
where $\,\varepsilon_{ij}^E\,$ and $\,\varepsilon_{ij}^{T_2}\,$ are the components of the
tensor of the $\,E\,$- and $\,T_2\,$-deformations of the octahedron in the $\,abc\,$-
system, respectively.

Proceeding to the permittivity tensor $\,\hat \epsilon\,$ and summing over all
Fe-ions sites, we arrive at nonzero diagonal components of $\,\hat \epsilon\,$:
\begin{equation}
\epsilon_{ii}\>=\>P_E \varepsilon_{ii}^E \>+\>P_{T_2}\varepsilon_{ii}^{T_2}\>,\label{eq:67}
\end{equation}
where $\,P_{E,T_2}\,=\,4\pi N \Bigl(\frac{n_0^2\,+\,2}{3}\Bigr)^2p_{E,T_2}\,$;
$\,\>N\,$ is the number of Fe$^{3+}$ ions per 1 $\,cm^3\,$. Components of
$\,\hat \varepsilon^E,\> \hat \varepsilon^{T_2}\,$ tensors serve as the {\it structure factors}
and may be calculated taking into account the known components of the tensor of
FeO$_6$ octahedron local deformations and the Eulerian angles relating the local axes to the $\,abc\,$ ones.

Thus, we have a two-parameter formula
(\ref{eq:67}) for the birefringence of orthoferrites as a function of rhombic
distortions of their crystal structure.
The photoelastic constants $\,P_E,\> P_{T_2}\,$
can be found from the comparison of experimental data \cite{Clover,Tabor} with
the theoretical structure dependence of the $\,ab\,$-plane birefringence :
\begin{equation}
\Delta n_{ab}\,=\,n_{a}-n_{b}\,=\,\frac{1}{2n_0}\left[P_E(\varepsilon_{xx}^E\,-\,\varepsilon_{yy}^E)\,+\,
P_{T_2}(\varepsilon_{xx}^{T_2}\,-\,\varepsilon_{yy}^{T_2})\right]    \label{eq:68}
\end{equation}
treated as a dependence on the type of the orthoferrite. The Figure\,\ref{fig8} shows both
experimental and calculated $\,\Delta n_{ab}\,$  given $\,P_E\,=\,6.2\,n_0\,,
\>P_{T_2}\,=\,4.0\,n_0\,$ (values obtained from the least-squares fitting).
A cogent agreement of the two-parameter formula (\ref{eq:68}) with experiment
testifies to the validity of the deformation model of the birefringence.

\begin{figure}[htbp]
	\centering
		\includegraphics[width=8cm,angle=0]{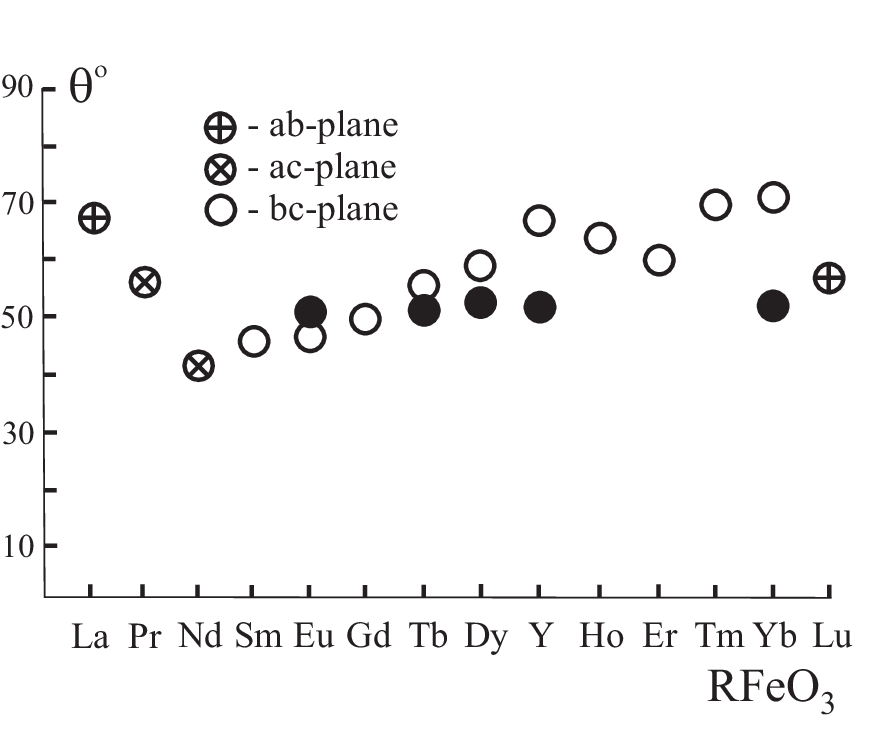}
		\caption{\footnotesize
		The orientation angles ($\pm\theta$) of optical axes in respective planes of orthoferrites predicted by the deformation model. The solid black circles are scarce experimental data for $bc$-plane (see text for detail).}
	\label{fig9}
\end{figure}

Using the found parameter $\,P_{E,T_2}\,$  values, we are able to describe all the
peculiarities of the orthoferrite birefringence. In particular, Fig.\,\ref{fig9} shows the theoretical predictions for the orientation angles $\pm\theta$ of optical axes, measured from the $c$-axis for the $ac$- and $bc$-planes and from the $a$-axis for the $ab$-plane, together with  scarce
experimental data on Eu,\,Tb,\,Dy,\,Y,\,Yb orthoferrites\,\cite{Tabor,Chetkindi}. Quite good agreement with the available experimental data is another confirmation of the validity of the deformation model of birefringence of orthoferrites. In general, for all its simplicity, the deformation model reflects quite correctly the main peculiarities of the natural birefringence of orthoferrites.
Moreover, the deformation model enables us to analyze the photoelastic  effects in orthoferrites.

\subsection{Photoelastic effects in orthoferrites}

 The elastic state of a lattice  is characterized by the macroscopic deformations
tensor $\,\hat \varepsilon\,$ and by the sublattice displacements not related to a change
of macroscopic crystal sizes, so-called {\it hidden displacements}.
The importance of hidden displacements was pointed out in Refs.
\cite{Jahn,Belanger}.

Within a linear approximation the permittivity tensor related to macroscopic deformations and hidden displacements of Bravais sublattices as follows:
\begin{equation}
\epsilon_{ij}\,=\,\epsilon_{ij}^o\,+\,
P_{ijkl}^o\varepsilon_{kl}\,+\,P_{ij}^n(\Gamma_{\nu})u_n(\Gamma_{\nu})
\>, \label{eq:69}
\end{equation}
where $\,P_{ijkl}^o\,$ and $\,P_{ij}^n\,$ are tensors of the photoelastic
constants, $\> \epsilon_{ij}^o\,$ is the
permittivity tensor in absence of deformations and displacements.

To compute $\,P_{ijkl}^o\,$ and $\,P_{ij}^n\,$, one can use the deformation
model of birefringence. The procedure is as follows :

1) local $\,E\,$- and $\,T_2\,$-type deformations of the FeO$_6$
complex as functions of the macrodeformations $\,\hat \varepsilon\,$ and displacements
$\,u_n\,$ are to be found ;

2) obtained $\, \varepsilon_{ij}^E\,$ and $\, \varepsilon_{ij}^{T_2}\,$ values pertaining to the
local octahedron axes are to be recalculated for the $\,abc\,$-system and
substituted in (\ref{eq:67}) ;

3) the resulting linear relation of $\, \epsilon_{ij}\,$ to macrodeformations
and displacements must be compared with (\ref{eq:69}), all photoelastic constants
$\,P_{ijkl}^o\,$ and $\,P_{ij}^n\,$ as functions of two parameters -- $\,P_E\,$
and $\,P_{T_2}\,$ -- being hereby determined.

The relation of the tensor of FeO$_6$ complexes microdeformations
to displacements of $\,O^{2-}\,$- ions is given by Exp.\,(\ref{u}):

Photoelastic constants $\,P_{ijkl}^0\,$ for YFeO$_3$, calculated in the way
described above are given in Ref.\,\cite{Molagu}. Assuming that the photoelastic constants for NdFeO$_3$
and YFeO$_3$ are close in magnitude and vary slightly with temperature, the authors
have evaluated the change of permittivity tensor components for NdFeO$_3$
as the temperature lowers from 293\,K to 8\,K. Data of neutron diffraction
study of the NdFeO$_3$ crystal structure\,\cite{Sosnowska} and values of magnetoelastic
constants for YFeO$_3$\,\cite{Molagu} have been used. The authors have obtained the
following values :$\quad \Delta\epsilon_{xx}\,=\,-(0.1\,- \,0.6)\cdot 10^{-3}
\>n_0\,;\quad \Delta\epsilon_{yy}\,=\,+(8.5\,+\,8.6)\cdot 10^{-3}\>n_0\,;
\quad \Delta\epsilon_{zz}\,=\,-(8.4\,+\,9.3)\cdot 10^{-3}\>n_0\,$.
Here, the first term is due to the macroscopic deformations, and the second
term is the contribution of hidden displacements of $\,O^{2-}\,$ ions .
Note that in all $\,\hat \epsilon\,$ components, the second term is larger than the
first one, i.e., one may not neglect the hidden displacements contribution.

When the direct action (external with respect to the elastic subsystem) on the
hidden displacements is lacking, i.e., in the free energy of the harmonic
crystal there exist no terms linear in $\,u_n\,$, the hidden displacements
are related to macrodeformations (see Exp.\,(\ref{A})). In this case, the third term in (\ref{eq:69}) can be reduced to the
second term, the
photoelastic constants $\,P_{ijkl}\,$ being thereby renormalized :
\begin{equation}
P_{ijkl}\,=\,P_{ijkl}^0\,+\,\Delta P_{ijkl}\,;\quad \Delta P_{ijkl}\,=\,
\sum_{n}P_{ij}^n(\Gamma_{\nu})A_{kl}^n(\Gamma_{\nu})\,.\label{eq:72}
\end{equation}
The values of $\, \Delta P_{ijkl}\,$ for YFeO$_3$
were estimated in Ref.\,\cite{Molagu} with making use of the results of the model calculation of inner stress tensor components
for TmFeO$_3$. The hidden displacements make an
appreciable, and sometimes leading, contribution to the photoelastic constants. 

\subsection{Photomagnetoelastic effects in orthoferrites}

Minimizing $\,\Phi_{me}\,+\,\Phi_e\,$ (see Exps. (\ref{Pme}) and (\ref{Pe}))  in $\,\hat \varepsilon\,$ and $\,u_n\,$, one can
determine their equilibrium values. Substituting
these values in (\ref{eq:69}), we obtain the basic formula for the analysis of
photomagnetoelastic effects \cite{Molagu} :
\begin{eqnarray}
\epsilon_{ij}\,=\,\epsilon_{ij}^0\,+\,
P_{ijkl}\varepsilon_{kl}\,+\,
Q_{ij \alpha \beta}G_{\alpha}\,G_{\beta}  \>,\nonumber \\
  Q_{ij \alpha \beta}\,=\,-\,\sum_{\Gamma_{\nu}mn}P_{ij}^n(\Gamma_{\nu})
\left[C^{-1}(\Gamma_{\nu})\right]^{nm}\Pi_{\alpha \beta}^m(\Gamma_{\nu})\,.\label{eq:75}
\end{eqnarray}
According  to this formula, the photomagnetoelastic effect includes two terms :
the first one, purely $\, magnetostrictive\,$ in nature, is defined by ordinary
photoelastic constants $\,P_{ijkl}\,$; another term is due to the magnetoelastic
displacements \cite{Molagu} and does not depend on the magnetostrictive deformations.
In other words, {\it there may exist the photomagnetoelastic effect even if the
magnetostriction is absent}!  
It is worth noting that, in fact, the Exp.\,\ref{eq:75} describes one of the mechanisms of quadratic magnetooptic Cotton-Mouton effect.

The quantitative evaluation of the
photomagnetoelastic constants $\,Q_{ij \alpha \beta}\,$ is rather complicated. The constants were evaluated in Ref.\,\cite{Molagu}
 using the data of the model calculations of the elastic
parameters $\,C^{nm}\,$, the magnetoelastic parameters
$\,B_{\alpha \beta}^m\,$, and the photoelastic parameters
$\,P_{ij}^n\,$. 
To show the actual significance of the magnetoelastooptic effect in orthoferrites,
the authors\,\cite{Molagu} have calculated the birefringence change in YFeO$_3$ at the
spin-reorientation $\,\Gamma_4\,\rightarrow \,\Gamma_2\quad(G_x\, \rightarrow
\,G_z)\,$  induced by the external magnetic field (${\bf H}\parallel a$-axis). The spin-reorientation is completed at $\,H\,=\,$75
$\,kOe\,$ and accompanied with magnetostrictive deformations $\,\varepsilon_{ii}\,=\,
\varepsilon_{ii}^0\sin^2\theta\,,\quad \varepsilon_{xz}\,=\,\varepsilon_{xz}^0\sin\,2\theta\,$, where
$\,\varepsilon_{xx}^0\,=\,\varepsilon_{yy}^0\,=\,1.8\cdot 10^{-5},\> \varepsilon_{zz}^0\,=\,
-\,3.7\cdot 10^{-5}\,,\>|\varepsilon_{xz}^0|\,=\,0.3\cdot 10^{-5}\,$
\cite{KP,JETPLett-1981}.
\begin{center}
\begin{table}
\caption{Different contributions to the change of birefringeance at the $\Gamma_4 - \Gamma_2$ spin-reorientation in YFeO$_3$ ($\lambda$\,=\,0.63\,$\mu$m, $T$\,=\,300\,K, $\Delta n_{ij}=\Delta (n_i-n_j)$).}
\centering
\begin{tabular}{|c|c|c|c|c|c|c|}
\hline
 $\Delta n_{ij}, \epsilon_{xz}^{(0)}$ ($\times 10^{-4}$)  &  $\Delta n_{ab}$ &  $\Delta n_{ac}$ &  $\Delta n_{bc}$ &$\epsilon_{xz}^{(0)}$  \\ \hline
  Magnetostriction &  -0.1 & 1.7 & 1.8 & $\pm$0.3 \\ \hline
  Hidden displacements & -0.7
  & 0.1 &  0.8 & $\pm$0.7 \\ \hline
 Total &-0.8
& 1.8 & 2.6 & $\pm$1.0 ($\pm$0.4)\\ \hline
 Experiment\,\cite{Kripiru} & -0.6$\pm$0.4
& 2.0$\pm$0.2 & 2.6$\pm$0.2 &  $\pm$0.7\\ \hline
 \end{tabular}
\label{PMEO}
\end{table}
\end{center}

Table\,\ref{PMEO} shows the estimations of the purely magnetostrictive contribution,
the magnetoelastic displacements contribution, as well as the total 
contribution to $\quad \Delta n_{ab}\,,\> \Delta n_{bc}\,,\>\Delta n_{ac}\>$
$\Bigl($we denote $\>\Delta n_{ij}\,=\,\Delta \,(n_i\,-\,n_j) \Bigr)\,$) as well
as the maximal $\,\epsilon_{xz}\,$ value  $\epsilon_{xz}^0\,$ at the $\,\Gamma_4\,\rightarrow \,
\Gamma_2\,$ transition in  YFeO$_3$ ($\lambda$\,=\,0.63\,$\mu$m\,, $T$\,=\,300\,K) as compared with experimental data\,\cite{Kripiru}. Note that two magnetoelastic terms are comparable in
magnitude, and so, both mechanisms of forming the orthoferrite birefringence
should be taken into account.  The theoretical predictions for the total
magnetoelastic contribution to the birefringence change at the $\,\Gamma_4\,
\rightarrow \,\Gamma_2\,$ transition reasonably agree with the experimental
data enabling us to draw the conclusion that the photomagnetoelastic effects
are dominant in forming the magnetic birefringence for the yttrium orthoferrite YFeO$_3$.

Thus, a simple deformation model based on the relation of the FeO$_6$
octahedron polarization to its deformation permits to explain the observed
peculiarities of the natural birefringence of orthoferrites and to calculate
all photoelastic and photomagnetoelastic effects in orthoferrites.
Thus, besides the lattice macrodeformations, an important role in photoelasto- and
photomagnetoelastic effects belongs to the hidden displacements of sublattices.

The analysis made can be extended to other compounds, too.  Always one should
keep in mind that the information about the photoelastic constants $\,P_{ijkl}\,$
 and macrodeformations $\,\hat \varepsilon\,$ of the crystal is, in general,
 $\,insufficient\,$ to consider the photoelastic and, especially,
photomagnetoelastic effects. Indeed, the same macrodeformation ensuing from  the external mechanical stress application;  the temperature change with/without the magnetic order alteration; the application of the external magnetic field 
can be accompanied with different hidden displacements of sublattices resulting
in different birefringence. These circumstances may seemingly be a cause of the
opposite thermal and pressure behaviour of the MnF$_2$ birefringence\,\cite{Markovin} at the same macrodeformation.

\section{Summary}

We applied simple physically clear theoretical approach to evaluate the interplay between FeO$_6$ octahedral distortions/rotations in rare-earth orthoferrites 
 and main magnetic and optic characteristics such as superexchange integral and N\'{e}el temperature, overt and hidden canting of magnetic sublattices, magnetic and magnetoelastic anisotropy, optic and photoelastic anisotropy. Most attention in the paper focused on the  Dzyaloshinskii vector, its value, orientation, and sense. Our analysis once again confirms the unambiguous leading role of antisymmetric exchange in the formation of overt and hidden canting in orthoferrites and the fallacy of the argumentation of the authors of the recent paper\,\cite{Zhou}.

The model approach developed in this work goes far beyond the scope of only orthoferrites.
Our analysis revealed previously underestimated relationships that can be used not only to elucidate the mechanisms of formation of various physical properties, but also to design electronic
structures for advanced materials.
Importantly, the relationships established with these model approaches may be cross-validated by the construction of hybrid data sets, which combine theoretical results
with experiment data, making it possible to extract and
validate new insight into the material physics of strongly correlated oxides\,\cite{str-pro}. We anticipate that this approach will spawn a number of additional studies for perovskites and other crystals  since it is immediately generalizable: the synergy of simple cluster models with subsequent first-principles calculations provides a platform to achieve rational structure-driven design of complex materials. A good understanding of the structure-property relationships can be used to develop new functional materials and devices.



\begin{acknowledgments}
I thank E.V. Sinitsyn and I.G. Bostrem for very fruitful multi-year collaboration, I.E. Dzyaloshinskii, V.I. Ozhogin, R.E. Walstedt, S.V. Maleev, B.Z. Malkin, B.S. Tsukerblatt, S.-L. Drechsler, and V.E. Dmitrienko for stimulating and encouraging discussions. This study was supported by the Ministry of Science and Higher Education of the Russian Federation, project FEUZ-2023-0017.
\end{acknowledgments}



\section*{References}

\begin{thebibliography}{199}


\bibitem{Geller}
S. Geller and E.~A. Wood, Acta Cryst. {\bf 9}, 563 (1956); J.Chem.Phys. {\bf 24}, 1236 (1956).

\bibitem{SFO}
 Ahmed, Shahran; Nishat, Sadiq Shahriyar; Kabir, Alamgir; {\it et al.}
PHYSICA B-CONDENSED MATTER  {\bf 615},    413061   (2021).

\bibitem{CM-2019}
A.~S. Moskvin, Condens. Matter. {\bf 4}(4), 84 (2019).


\bibitem{str-pro}
Prasanna V. Balachandran and James M. Rondinelli, Phys. Rev. B {\bf 88}, 054101 (2013).

\bibitem{Olekh}
N.~M. Olekhnovich, Crystallography Reports, {\bf 52}, 759–767 (2007).

\bibitem{Zhou}
J.-S. Zhou, L.~G. Marshall, Z.-Y. Li, X. Li, and J.-M. He,
Phys. Rev. B {\bf 102}, 104420 (2020).

\bibitem{muon}
 E. Holzschuh, A.~B. Denison, W. Kundig, P.~F. Meier, and B.~D. Patterson, Phys. Rev. B {\bf 27} 5294 (1983).


\bibitem{Amelin}
K. Amelin, U. Nagel, R.~S. Fishman, Y. Yoshida, Hasung Sim, Kisoo Park, Je-Geun Park, and T. R\~o\~om
Phys. Rev. B {\bf 98}, 174417 (2018).

\bibitem{KP}
K.~P. Belov, A.~K. Zvezdin, A.~M. Kadomtseva, and R.~Z. Levitin, Orientational
Transitions in Rare-Earth Magnetics, Nauka, Moscow 1979 (in Russian).

\bibitem{1971}
A.~S. Moskvin, Fizika Tverdogo Tela, {\bf 12}, 3209 (1970) (Soviet Physics Solid State, USSR {\bf 12}, 2593 (1971)).

\bibitem{thesis}
A.~S. Moskvin, Antisimmetrichniy obmen i magnitnaya anizotropiya v slabyh ferromagnetikah (Antisymmetric exchange and magnetic anisotropy in weak ferromagnets). D.Sc  Dissertation, Lomonosov Moscow  State University, 1984 (in Russian).

\bibitem{Sidorov}
A.~A. Sidorov, A.~S. Moskvin, V.~V. Popkov, Fizika Tverdogo Tela, {\bf 18}, 3005 (1976).

\bibitem{Ovanesyan}
A.~S. Moskvin, N.~S. Ovanesyan, and V.~A. Trukhtanov, Hyperfine Interactions, {\bf 1}, 265 (1975).

\bibitem{JMMM-2016}
A.~S. Moskvin, JMMM, {\bf 400}, 117, (2016).




\bibitem{JETP-2021}
A.~S. Moskvin, JETP, {\bf 132}, 517–547 (2021).

\bibitem{Freeman}
S. Freeman, Phys. Rev. B {\bf 7}, 3960 (1973).

\bibitem{Luk}
A.~S. Moskvin and A.~S. Luk'yanov,  Sov. Phys. Solid State {\bf 19}, 701 (1977).

\bibitem{LuCrO3}
R.~M. Hornreich, S. Shtrikman, B.~M. Wanklyn, and I. Yaeger, Phys. Rev. B {\bf 13}, 4046 (1976).

\bibitem{Bloch}
D. Bloch, J. Phys. Chem. Solids, {\bf 27}, 881 (1966).

\bibitem{Dzyaloshinskii}
I.~E. Dzyaloshinskii, Soviet Physics JETP, {\bf 5} 1259 (1957); I.Dzyaloshinsky, J. Phys. Chem. Solids {\bf 4}, 241 (1958).


\bibitem{Moriya}
T. Moriya, Phys. Rev. Lett. {\bf 4}, 228 (1960); Phys. Rev. {\bf 120}, 91 (1960).

\bibitem{Keffer}
F. Keffer, Phys. Rev. {\bf 126}, 896 (1962).

\bibitem{Herrmann}
G.~F. Herrmann, Phys. Rev. {\bf 133}, A1334 (1964).


\bibitem{1977}
A.~S. Moskvin and I.~G. Bostrem, Fizika Tverdogo Tela, {\bf 19}, 2616 (1977) [Sov. Phys. Solid State {\bf 19}, 1532 (1977)].

\bibitem{Tofield}
B.~C. Tofield, B.~F.~E. Fender, J. Phys. Chem. Solids, {\bf 31}, 2741 (1970).

\bibitem{JMMM-2018}
A.~S. Moskvin, JMMM, {\bf 463}, 50-56 (2018).

\bibitem{RFeO3}
 M. Marezio, J.~P. Remeika, and P.~D. Dernier, Acta Crystallogr.,
Sect. B: Struct. Sci., Cryst. Eng. Mater. {\bf 26}, 2008 (1970); M. Marezio and P.~D. Dernier, Mater. Res. Bull. {\bf 6}, 23 (1971).

\bibitem{1975}
A.~S. Moskvin, E.~V. Sinitsyn, Fizika Tverdogo Tela, {\bf 17}, 2495 (1975) [Soviet Physics Solid State, {\bf 17}, 2495 (1975)].

\bibitem{Jacobs}
S. Jacobs, H.~F. Burne, and L.~M. Levinson, J. Appl. Phys. {\bf 42}, 1631 (1971).


\bibitem{Luetgemeier}
H. Luetgemeier, H.~G. Bohn and M. Brajczewska, J. Magn. Magn. Mat., {\bf 21}, 289 (1980).

\bibitem{Plakhtii}
V.~P. Plakhtii, Yu.~P. Chernenkov, J. Schweizer, and M.~N. Bedrizova, JETP, {\bf 53}, 1291 (1981); V.~P. Plakhtii, Yu.~P. Chernenkov, M.~N. Bedrizova, and J. Schweizer, AIP Conference Proceedings, {\bf 89} 330 (1982); V.~P. Plakhtii, Yu.~P. Chernenkov, M.~N. Bedrizova, Solid State Commun. {\bf 47}, 309 (1983).
\bibitem{Georgieva}
D.~G. Georgieva, K.~A. Krezhov, and V.~V. Nietza, Solid State Commun. {\bf 96}, 535 (1995).
\bibitem{sign}
A.~S. Moskvin,  Fizika Tverdogo Tela, {\bf 32}, 1644 (1990) [Sov. Phys. Solid State {\bf 32}, 959 (1990)].

\bibitem{Dmitrienko}
V.~E. Dmitrienko, E.~N. Ovchinnikova,	S.~P. Collins,  G. Nisbet,  G. Beutier, Y.~O. Kvashnin, V.~V. Mazurenko,	A.~I. Lichtenstein, and M.~I. Katsnelson, Nature Physics  {\bf 10}, 202 (2014).




\bibitem{DyFeO3}
A.~V. Zalesskii, A.~M. Savvinov, I.~S. Zheludev, A.~N. lvashchenko, JETP. {\bf 41}, 723 (1975).

\bibitem{DM}
A.~S. Moskvin, JETP, {\bf 104}, 911–925 (2007).

\bibitem{md}
A.~S. Moskvin, E.~V. Sinitsyn, and A.~Yu. Smirnov,  Sov. Phys. Solid State {\bf 20}, 2002 (1978).

\bibitem{TIA}
A.~S. Moskvin,  I.~G. Bostrem, M.~A. Sidorov,  JETP, {\bf 77}, 127 (1993).

\bibitem{Volkov}
A.~A. Volkov, Yu.~G. Goncharov,  G.~V. Kozlov,  K.~N. Kocharyan,  S.~P. Lebedev,  A.~S. Prokhorov,  A.~M. Prokhorov,  JETPLett, {\bf 39}, 166 (1984).
\bibitem{White-1982}
R.~M. White, R.~J. Nemanich, and C. Herring, Phys. Rev. B {\bf 25}, 1822 (1982).

\bibitem{Hahn}
S.~E. Hahn, A.~A. Podlesnyak, G. Ehlers, G.~E. Granroth, R.~S. Fishman, A.~I. Kolesnikov, E. Pomjakushina, and K. Conder, Phys. Rev. B {\bf 89}, 014420 (2014).

\bibitem{Park}
K. Park, H. Sim, J.~C. Leiner, Y. Yoshida, J. Jeong, S. Yano,
J. Gardner, P. Bourges, M. Klicpera, V. Sechovský, M. Boehm,
and J.-G. Park, J. Phys.: Condens. Matter {\bf 30}, 235802 (2018).

\bibitem{Licht}	
 A.~I. Likhtenshtein, A.~S. Moskvin, and V.~A. Gubanov, Fiz. Tverd. Tela
{\bf 24}, 3596 (1982) (Soviet Phys. Solid State {\bf 24}, 2049 (1982)).

 \bibitem{JETP-1981}
A.~M. Kadomtseva, A.~P. Agafonov, M.~M. Lukina, V.~N. Milov, A.~S. Moskvin, V.~A. Semenov, E.~V. Sinitsyn, JETP, {\bf 81}, 700-706 (1981).

\bibitem{Bidaux}
R. Bidaux, S.E. Bouree, J.~J. Hammann, Phys. Chem. Solids. {\bf 35}. P. 1645-1655 (1974). 

\bibitem{Mukhin}
 A.~A. Egoyan, A.~A. Mukhin, Fizika Tverdogo Tela {\bf 36},   1715-1723 (1994).

\bibitem{JETPLett-1981}
A.~M. Kadomtseva, A.~P. Agafonov, V.~N. Milov, A.~S. Moskvin, and V.~A. Semenov,
Zh. eksper. teor. Fiz., Pisma {\bf 33}, 400 (1981).

\bibitem{FTT-1987}
A.~S. Moskvin, D.~G. Latypov, A.~P. Agafonov, Fizika Tverdogo Tela, {\bf 29}, 3157-3160 (1987) (Sov. Phys. Solid State {\bf 29}, 1814 (1987)).


\bibitem{Cullen}
James R. Cullen and Arthur E. Clark, Phys.Rev.B {\bf 15}, 4510 (1977).

\bibitem{Bumagina}
L.~A. Bumagina, V.~I. Krotov, B.~Z. Malkin, A.~K. Khasanov,  Zhurnal Eksperimentalnoi I Teoreticheskoi Fiziki, {\bf 80}, 1543-1553 (1981).

\bibitem{cubic}
A.~S. Moskvin and I.~G. Bostrem, Sov. Phys. Solid St. {\bf 21}, 628 (1979).


\bibitem{Kahn}
F.~J. Kahn, P.~S. Pershan, and J.~P. Remeika,  Phys. Rev. {\bf 186}, 891 (1969).

\bibitem{Z1}
A.~S. Moskvin, A.~V. Zenkov, E.~I. Yuryeva, and V.~A. Gubanov, Physica B {\bf 168}, 187 (1991).

\bibitem{Z2}
A.~S. Moskvin, A.~V. Zenkov, E.~A. Ganshina, G.~S. Krinchik, and M.~M. Nishanova,
J.Phys.Chem.Solids {\bf 54}, 101 (1993).

\bibitem{Clover}
R.~B. Clover, C. Wentworth, and S.~S. Mroczkowski, IEEE Trans. Magn.
{\bf 7}, 480 (1971).

\bibitem{Tabor}
W.~J. Tabor, A.~W. Anderson, and L.~G. van Uitert,  J. Appl. Phys. {\bf 41} (7), 3018 (1970).

\bibitem{Chetkindi}
M.~V. Chetkin, Yu.~S. Didosyan, and A.~I. Akhutkina, Fiz. Tverd. Tela {\bf 13}, 3414
(1971).





\bibitem{Clover1}
R.~B. Clover, M. Rayl, and D. Gutman, in: Magn. and Magn. Mater., 17th AIP Annu.
Conf. Chicago 1971, New York, 1972 (Part 1, p.264).

\bibitem{Jahn}
I.~R. Jahn, phys.stat.sol. (b) {\bf 57}, 681 (1973).

\bibitem{Belanger}
D.~P. Belanger, A.~R. King, and Y. Jaccarino, Phys.Rev.B {\bf 29}, 2636 (1984).

\bibitem{Molagu}
A.~S. Moskvin, D.~G. Latypov, and V.~G. Gudkov, Fiz. Tverd. Tela {\bf 30}, 413
(1988).

\bibitem{Sosnowska}
I. Sosnowska and  P. Fischer, in: Neutron Scattering Symp. Argonne 1981,
New York, 1982 (p.346).



\bibitem{Kripiru}
B.~B. Krichevtsov, R.~V. Pisarev, and M.~M. Ruvinshtein, Fiz. Tverd. Tela
{\bf 22}, 2128 (1980).

\bibitem{Markovin}
P.~A. Markovin and R.~V. Pisarev, Zh. eksper. teor. Fiz. {\bf 77}, 2461 (1979).


\end{thebibliography}
\end{document}